\pdfoutput=1

\documentclass[%
reprint,
superscriptaddress,
nofootinbib,
nobibnotes,
 amsmath,amssymb,
 aps,
pra,
nofootinbib,
nobibnotes,
]{revtex4-1}

\usepackage{graphicx}
\usepackage{dcolumn}
\usepackage{bm}

\usepackage{subfig}

\usepackage{amsmath}
\usepackage{mathtools} 

\usepackage{xcolor}

\usepackage{hyperref}
\hypersetup{
	colorlinks=true,
    linkcolor=blue,
    filecolor=blue,      
    urlcolor=blue,
    citecolor=blue,
    pdftitle={Draft_OC_NLSMs_Munoz-Segovia_and_Cortijo},    
    pdfauthor={Daniel Muñoz-Segovia},     
    pdfkeywords={nodal-line semimetals, optical conductivity, Coulomb interaction}, 
}
 
\begin{document}

\title{Many-body effects in nodal-line semimetals: Correction to the optical conductivity}

\author{Daniel Mu\~noz-Segovia}%
\email{daniel.munozsegovia@dipc.org}
\affiliation{Instituto de Ciencia de Materiales de Madrid, CSIC, Cantoblanco, 28049 Madrid, Spain}%
\affiliation{Donostia International Physics Center, 20018 Donostia-San Sebastian, Spain}%
\author{Alberto Cortijo}%
\email{alberto.cortijo@uam.es}
\affiliation{Instituto de Ciencia de Materiales de Madrid, CSIC, Cantoblanco, 28049 Madrid, Spain}
\affiliation{Departamento de F\'isica de la Materia Condensada, Universidad Aut\'onoma de Madrid, Madrid E-28049, Spain}

\date{\today}

\begin{abstract}
Coulomb interaction might have important effects on the physical observables in topological semimetals with vanishing density of states at the band touching due to the weak screening. In this work, we show that Kohn's theorem is not fulfilled in nodal-line semimetals (NLSMs), which implies non-vanishing interaction corrections to the conductivity. Using renormalized perturbation theory, we determine the first-order optical conductivity in a clean NLSM to be $\sigma_{\perp \perp}(\Omega) = 2 \sigma_{\parallel \parallel}(\Omega) = \sigma_0 [1 + C_2 \alpha_R(\Omega)]$, where $\perp$ and $\parallel$ denote the perpendicular and parallel components with respect to the nodal loop, $\sigma_0 = (2 \pi k_0) e^2/(16h)$ is the conductivity in the noninteracting limit, $2 \pi k_0$ is the nodal loop perimeter, $C_2 = (19-6\pi)/12 \simeq 0.013$ is a numerical constant and $\alpha_R(\Omega)$ is the renormalized fine structure constant in the NLSM. The analogies between three-dimensional NLSMs and two-dimensional Dirac fermions are reflected in the parallelism between their respective optical conductivities, both in the noninteracting limit and in the correction, as pointed out by the equality of the universal coefficient $C_2$ in both systems. Finally, we analyze some experiments that have determined the optical conductivity in NLSMs, discussing the possibility of experimentally measuring our result.


\end{abstract}

\maketitle

\section{\label{section:Introduction} Introduction}




Modern condensed matter physics has an increasing interest in the study of systems based on their topological properties. While the most renowned classes of topologically nontrivial materials are the topological insulators \cite{Hasan2010}, which are gapped in the bulk but have protected gapless surface states, more recent work has shown the existence of topologically nontrivial materials which are also gapless in the bulk, the so-called topological nodal semimetals \cite{Burkov2016,Weng2016,Armitage2018}. These materials are characterized by a crossing between the conduction and valence bands closest to the Fermi level. While this crossing is protected by certain symmetries (i.e., it cannot be removed by symmetry-preserving perturbations), its robustness depends crucially on the codimension of the band-touching node, i.e., on the difference between the spatial dimension and the defect dimension. In three dimensions (3D), two cases must be differentiated. Weyl \cite{Murakami2007,Wan2011} and Dirac \cite{Wang2012} semimetals, in which the nodes consist of zero-dimensional (0D) discrete nodal points, are the most robust variety. For instance, the presence of either time-reversal symmetry or inversion symmetry (not both simultaneously) guarantees the topological stability of Weyl semimetals. 
The crossing may also be a one-dimensional (1D) nodal line \cite{Burkov2011}, 
either twofold degenerate (Weyl type) or fourfold degenerate (Dirac type). Nodal lines might appear in different shapes: extended lines running across the Brillouin zone (BZ) 
\cite{Shao2019}, closed loops \cite{Burkov2011}, chains of loops \cite{Bzdusek2016}, linked rings \cite{Yan2017,Chang2017}, knotted loops \cite{Bi2017}, etc. 

This line node defines the so-called nodal-line semimetals (NLSMs), which are the focus of this work. Whereas NLSMs do not posses the robustness of point nodes, the presence of some additional symmetries 
can stabilize them, and indeed depending on the protecting symmetries different topological invariants can be defined \cite{Fang2016a,Chiu2014,Zhao2016,Chan2016}. For instance, two $\mathbb{Z}_2$ invariants have been found when time-reversal, inversion and spin-rotation (i.e., absence of spin-orbit coupling) symmetries apply \cite{Fang2015,Kim2015}, while stability when spin-rotation symmetry is broken requires additional (nonsymmorphic) symmetries to be imposed, such as a glide or twofold screw symmetry \cite{Armitage2018,Bian2016a,Chen2016}, and a $\mathbb{Z}$ index can be associated to the nodal line in this case. It has also been found that when a protecting symmetry is broken, the NLSM becomes either gapped or a Weyl or Dirac semimetal \cite{Chen2015,Carter2012,Weng2015}. 


However, in general, these conditions are insufficient to ensure the nodal line having constant energy \cite{Burkov2011}, and thus it is not generically located at the Fermi level. It is true though that a constant energy line may well be a good approximation, and indeed exact if particle-hole symmetry is present (as in nodal-line superconductors). Unlike most topological phases, NLSMs do not necessarily posses protected surface states \cite{Fang2015,Fang2016}, which would in general require the surfaces to preserve the symmetries that protect the line node. Nevertheless, even if this does not apply, when particle-hole symmetry (approximately) holds, a (nearly) flat, drumhead-like band appears over the surface BZ enclosed by the projection of the nodal line onto the corresponding surface \cite{Burkov2011}. However, the lack of topological protection for these surface states means that a change of the model parameters not necessarily breaking any particular symmetry might spoil their flatness and localization \cite{Fang2016a}.

Aside from their fundamental interest, NLSMs have also a practical interest due to their unusual transport properties \cite{Hu2019,Syzranov2017,Carbotte2017,Mukherjee2017,Barati2017,Rui2018}, and moreover they have been proposed as hydrogen catalysts due to their exotic surface states \cite{Li2018}. However, while there has been strong theoretical interest for NLSMs since their first proposal \cite{Burkov2011} and there exist a number of materials predicted to show these line nodes \cite{Weng2016}, experimental evidence has only appeared quite recently. To our knowledge there are a dozen solid-state materials in which nodal lines have been experimentally demonstrated, especially via angle-resolved photoemission spectroscopy (ARPES) \cite{Lv2019} or quantum magnetic oscillation measurements \cite{Oroszlany2018}. Among these materials, one can mention, for instance, ZrSiS \cite{Schoop2016,Neupane2016,Matusiak2017,Hu2017,Pezzini2018}, PbTaSe$_2$ \cite{Bian2016a,Xu2019} and CaAgAs \cite{Emmanouilidou2017,Takane2018a}. Let us highlight that, more lately, optical spectroscopy has emerged as a complementary experimental technique able to provide evidence to identify NLSMs \cite{Schilling2017,Shao2019,Qiu2019}. On the other hand, NLSMs have also been realized with cold atoms \cite{Song2019} as well as in photonic \cite{Gao2018,Yan2018,Xia2019} and mechanical \cite{Qiu2019a} metacrystals.






This work studies the effect of the long-range Coulomb interaction between electrons on a physical observable: the optical conductivity. The influence of the Coulomb interaction on the properties of other physical systems is well known. In Fermi liquid metals, although this long-range interaction is marginal in the renormalization group (RG) sense, the strong Thomas-Fermi screening arising from their extended Fermi surface with non-zero density of states (DOS) makes the Coulomb interaction effectively short-ranged (and thus irrelevant) \cite{Pines1994,Shankar1991,Houghton1994}. Physical properties in nodal point semimetals in both 2D (e.g. graphene) and 3D (e.g. Weyl semimetals) receive logarithmic corrections due to the Coulomb interaction remaining only marginally irrelevant \cite{Kotov2012,Isobe2012,barnes2014,Lai2015,Throckmorton2015} as a consequence of a vanishing DOS at the nodal points, which makes screening weak due to the few states available to participate. NLSMs also display a vanishing DOS at the line node \cite{Barati2017}, so the Coulomb interaction is expected to remain long-ranged and marginally irrelevant. Indeed, this has been found to be the case in clean NLSMs \cite{Huh2016}. Moreover, a nontrivial interacting fixed point at which the screened Coulomb interaction is irrelevant has been predicted by means of both RG and large-$N$ computations \cite{Huh2016}. This fact thus validates a perturbative treatment around this fixed point, despite the nominal ratio between kinetic and Coulomb energies being zero for vanishing chemical potential \cite{Rhim2016}. On the other hand, when disorder (by itself a marginally relevant perturbation) is taken into account, Coulomb interaction becomes marginally relevant by a feedback mechanism, although the Coulomb interaction flows to strong coupling asymptotically more slowly than disorder \cite{Wang2017}.



These considerations raise the question of the effect of the Coulomb interaction on the physical observables in NLSMs, which are expected to receive logarithmic corrections. This work focuses on the contribution of the Coulomb interaction to the optical conductivity, which being directly influenced by the charge dynamics thus provides information about the electronic band structure as well as about the correlations of the low-energy quasiparticles \cite{pavarini2016quantum}.

The corresponding problem of determining the Coulomb correction to the optical conductivity in graphene was the subject of a long controversy in the past. To first order in perturbation theory, the conductivity of intrinsic graphene at zero temperature reads as \cite{Herbut2008}:
\begin{equation}
\sigma_{gr}(\Omega) = \frac{e^2}{4 \hbar} \left[1+ C \alpha_{gr}(\Omega) +  \mathcal{O} \hspace{-2pt} \left(\alpha_{gr}^2\right)\right] ,
\end{equation}
where $\alpha_{gr}(\Omega) = e^2 / [4 \pi \varepsilon_0 \hbar v_{gr}(\Omega)]$ is the renormalized fine structure constant in suspended graphene, with $\varepsilon_0$ the vacuum permittivity and $v_{gr}(\Omega)$ the renormalized (physically measurable) Fermi velocity of graphene, and $C$ is a numerical constant. In particular, the controversy arose due to the fact that different calculations gave three different values for this first-order coefficient $C$ \cite{Teber2014}:
\begin{subequations}
\begin{align}
    & C_1 = \frac{25-6 \pi}{12} \simeq 0.512 , \\
	& C_2 = \frac{19-6 \pi}{12} \simeq 0.013 , \\
	& C_3 = \frac{22-6 \pi}{12} \simeq 0.263 .
\end{align}
\end{subequations}
What is more, the universality of this coefficient was also questioned, since it was proposed that high energy details not taken into account by the continuum Dirac model were necessary to correctly determine $C$.


It is worth noticing that experiments \cite{Nair2008, Kuzmenko2008} indicate that the overall effect of the Coulomb interaction is small. However, due to the fact that $\alpha_{gr} \sim 2$ in suspended graphene, the validity of perturbation theory, and in particular the smallness of the higher order terms in the perturbative expansion, is not guaranteed. Therefore, although $C_2$ is the value which better fits to experiments, one cannot rigorously conclude that this is correct result from phenomenology, but theoretical analysis is needed. On the other hand, experiments can be compared to non-perturbative calculations such as \cite{Boyda2016,Stauber2017}, both giving good agreement. 

There are mainly two important technical details in the calculation of $C$. Firstly, the initial approach for computing the conductivity from other quantity: from the density-density response, from the current-current correlator (Kubo formula), and from the kinetic quantum transport equation. Secondly, for calculations in the continuum Dirac model, the regularization procedure used: hard or soft cutoff, dimensional regularization (DR), implicit regularization, etc.




A thorough recapitulation of the theoretical calculations of $C$ can be found in \cite{Teber2018}. The calculation was first tackled in \cite{Herbut2008}, where the value $C_1$ was obtained using a hard-cutoff regularization with the Kubo approach. Shortly after, \cite{Mishchenko2008} concluded that in order for the three approaches, density-density response, Kubo and kinetic equation, to give a consistent result within a cutoff regularization, this has to be a soft one, in which case the three approaches provide the value $C_2$. This is a consequence of the fact that, in the approaches using the current operators (Kubo and kinetic equation), a na\"{i}ve hard-cutoff regularization violates the Ward identity, and thus gauge invariance. Nevertheless, a hard-cutoff regularization can also be used with these approaches as long as it is appropriately applied to enforce the Ward identity, as argued by \cite{Sheehy2009,Abedinpour2011} based on their Kubo hard-cutoff calculation. The density-density response approach does not suffer from this problem, and it gives the same result $C_2$ for all cutoffs \cite{Mishchenko2008, Sodemann2012}.

Then, \cite{Juricic2010} presented a DR calculation using both Kubo and density-density response approaches, both giving the $C_3$ result and respecting the Ward identity. However, as it is transparent in the more recent DR calculations of \cite{Teber2014,Teber2018}, DR gives the $C_2$ result using both the Kubo and density-density response approaches provided that one substracts \textit{all} the counterterms consistently order by order. The delicate point is that even if one does not substract the subdivergences in the first-order polarization tensor, a finite result is obtained, but one has to substract them in order to be consistent and properly renormalize the theory. The $C_3$ value was obtained again by \cite{Rosenstein2013a} using the Kubo approach and starting from a lattice model, but finally performing the integrals with a hard-cutoff without checking the Ward identity, which has the subtlety indicated above. Indeed, some of these authors obtained later $C_1$ using a similar method, and $C_2$ using the density-density response approach with a hard-cutoff \cite{Rosenstein2014}. At the same time, by applying implicit regularization to the Kubo approach and setting the arbitrariness by requiring the transversality of the polarization tensor, \cite{Gazzola2013} obtained $C_2$. Furthermore, a lattice tight-binding Kubo and a Wilson momentum-shell DR density-density response approach yielded $C_2$ \cite{Link2016}. Finally, an unscreened Hartree-Fock calculation \cite{Stauber2017} has obtained $C = 1/4 \simeq C_3$.

In summary, the more widely accepted conclusion is that the first-order coefficient $C$ is universal, i.e., independent of high-energy details and spatial extension of atomic wave functions \cite{Link2016}, and it has the small value $C_2$, which was first obtained by \cite{Mishchenko2008}, independently of the regularization scheme used. 




In this work, by mapping the NLSM problem to the graphene problem in the low-energy regime, we find that the optical conductivity of NLSMs obtains many-body corrections even at zero temperature and with the chemical potential pinned to the nodal line, analogously to graphene. It is worth highlighting that the low-energy region, where the NLSM displays a linear dispersion, corresponds to remaining at the leading-order term in a Taylor expansion in powers of the inverse of the nodal ring radius $1/k_0$, i.e., to the large-$k_0$ limit. Outside this regime, we expect frequency-dependent corrections $\mathcal{O}(\Omega/v_r k_0)$, both in the noninteracting limit, where they have already been determined in \cite{Barati2017}, and in the interaction correction. Evidently, these corrections would be model and thus material dependent, so that the analogy with graphene is no longer valid outside the linear regime. However, the study of the Coulomb interaction correction outside the linear regime, e.g. in lattice models, lies beyond the scope of this work, and we leave it for the future.

Within this linear dispersion model of the NLSM, the optical conductivity is anisotropic, with the component perpendicular to the nodal ring being twice the parallel ones: 
\begin{subequations}
\begin{align}
     & \sigma_{\perp \perp}(\Omega) = \sigma_0 \left[1 + C_2 \alpha_R(\Omega) + \mathcal{O}\hspace{-2pt}\left(\alpha_R^2\right)\right], \\
     & \sigma_{\parallel \parallel}(\Omega) = \frac{1}{2} \sigma_{\perp \perp}(\Omega),
\end{align}
\label{eq:conductivity_introduction}
\end{subequations}
\hspace{-3pt}where $\sigma_0=k_0 e^2 / (16 \hbar)$ is the noninteracting optical conductivity, $k_0$ is the radius of the nodal ring and $\alpha_R(\Omega) = e^2 / [4 \pi \varepsilon \hbar v_R(\Omega)] = \alpha_R(\bar{\mu}) / [1 + (1/4) \alpha_R(\bar{\mu}) \ln{(\bar{\mu} / \Omega)} ]$ is the renormalized fine structure constant in the NLSM, with $\varepsilon$ the static lattice dielectric constant of the NLSM, $v_R(\Omega)$ the renormalized (physically measurable) Fermi velocity and $\bar{\mu}$ the renormalization scale. Here, by means of dimensional regularization and both the Kubo and density-density response approaches (basically following the procedure of \cite{Teber2018}), we have obtained the value $C_2 = (19-6 \pi)/12 \simeq 0.013$ for the first-order coefficient for the NLSM, as determined by \cite{Teber2018} for graphene. As an aside, it is worth mentioning that, due to the NLSMs being 3D materials, it is the material's effective dielectric constant $\varepsilon$ that enters in the coupling constant $\alpha_R$ instead of the vacuum permittivity $\varepsilon_0$. Since the value of $\varepsilon$ might be of order of 10 (or more) times bigger than $\varepsilon_{0}$, the coupling $\alpha_R$ might now be much smaller than unity, and therefore perturbation theory is ensured to be well defined in NLSMs, in contrast to graphene.


As predicted by \cite{Burkov2011} based on the equal Fermi surface codimension (2) and DOS energy-dependence ($\propto|E-E_F|$), the optical conductivity of the NLSM shares analogies with the optical conductivity of graphene. In the noninteracting case, this fact was already discussed in Refs. \cite{Ahn2017,Rhim2016}, being the optical conductivity independent of the frequency and given by a universal value times the perimeter of the nodal ring ($2 \pi k_0$). This non-universal dependence comes from NLSMs living in 3D instead of 2D, and it significantly differentiates NLSMs also from Weyl semimetals, where the optical conductivity depends linearly with the frequency \cite{Roy2017}. Moreover, the analogies with graphene permeate to the interacting case. In the large-$k_0$ limit, the interaction correction to the optical conductivity is found to be exactly parallel to that of graphene (except for the factor $k_0$ in $\sigma_0$), so that it contains a logarithmic dependence on the frequency. It is also worth highlighting that, while this interaction correction $\sigma_0 C_2 \alpha_R(\Omega)$ is non-universal, its material dependence has exactly the same structure as that of the non-interacting conductivity (a proportionality to $k_0$), at least for sufficiently low frequency.

The paper is organized as follows. In Sec. \ref{model} we describe the model for a NLSM and the simplifications we will use to compute the corrections to the optical conductivity. Section \ref{sec:optical_conductivity} is devoted to the analysis of the optical conductivity (\ref{eq:conductivity_introduction}) of an interacting, clean NLSM. Finally, in Sec. \ref{section:Discussion} we discuss the features to consider in order to experimentally observe the previous result. Our result is compatible with the experimental uncertainties in current experiments measuring the optical conductivity, although there are no evidences of the logarithmic increase in frequency yet. Appendix \ref{appendix:Kohn} presents the proof of the failure of Kohn's theorem in NLSMs, which implies the appearance of non-vanishing interaction corrections to the conductivity. The technical details dealing with the derivation of the optical conductivity are provided in Appendix \ref{appendix:OC}.

\section{\label{model}NLSM model}

\begin{figure}[!t]
    \centering
    \subfloat{\includegraphics[width=.26\textwidth]{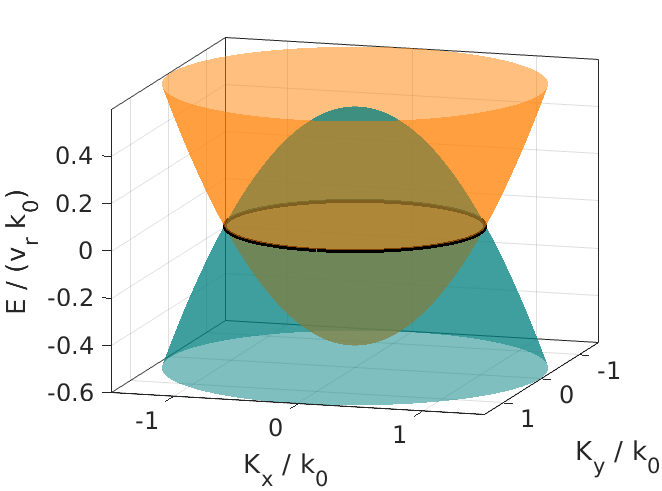}}
    \subfloat{\includegraphics[width=.23\textwidth]{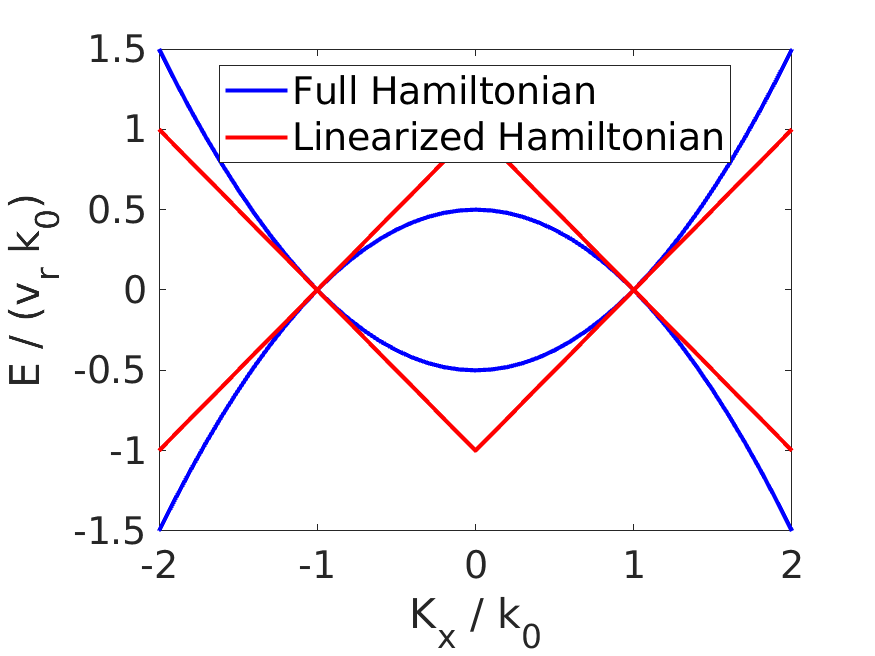}}
    \caption{Band structure of the NLSM with energy and momenta measured in units of $v_r k_0$ and $k_0$, respectively. Left: band structure of the full Hamiltonian (\ref{eq:first-quant-H0}) along the $K_x$ and $K_y$ directions for $K_z=0$. The line node is highlighted in black. Right: comparison of the band structures of the full (\ref{eq:first-quant-H0}) and linearized (\ref{eq:linear_Hamiltonian}) Hamiltonians along the $K_x$ direction for $K_y = K_z = 0$.}
    \label{fig:Band_structure}
\end{figure}

We begin by introducing the minimal continuum model of a NLSM we will use in this work. Following \cite{Fang2015,Chan2016,Huh2016,Oroszlany2018}, we consider two bands crossing each other in a circular loop in the $xy$-plane in momentum space, with the dispersion being parabolic in the $x$ and $y$ directions and linear in $z$. The second-quantized noninteracting Hamiltonian reads as $\hat{H}_0 = \sum_{\boldsymbol{K}} \hat{a}^{\dagger}_{\boldsymbol{K}} \mathcal{H}_0(\boldsymbol{K}) \hat{a}_{\boldsymbol{K}}$, with:
\begin{equation}
\mathcal{H}_0(\boldsymbol{K}) \hspace{-1pt} = \hspace{-2pt} \left( \Delta - \frac{K_x^2+K_y^2}{2 m} \right) \hspace{-1pt} \tau_x + v_z K_z \tau_z
\label{eq:first-quant-H0}
\end{equation}
where $\boldsymbol{K}$ is the canonical momentum and $\boldsymbol{\tau}$ are the Pauli matrices. 
The corresponding band structure, which is plotted in Fig. \ref{fig:Band_structure}, is then:
\begin{equation}
E_{\pm} (\boldsymbol{K}) = \pm \sqrt{\left( \Delta - \frac{K_x^2+K_y^2}{2 m} \right)^2 + v_z^2 K_z^2}.
\end{equation}
The bands touch each other at the circle defined by $K_x^2+K_y^2 = k_0^2$ in the $K_z = 0$ plane, where we have defined the nodal line radius $k_0 = \sqrt{2 m \Delta}$.

This model describes a rather general NLSM with time-reversal, inversion and reflection symmetries \cite{Chan2016}, as well as spin-rotation symmetry if the Pauli matrices act on the orbital degrees of freedom (in which case the line node would be of the Dirac type, with the additional degeneracy due to spin). This model accurately describes the low-energy dispersion relation of the NLSM candidate Ca$_3$P$_2$ \cite{Xie2015,Chan2016}, which displays negligible spin-orbit coupling as well as an almost energy-flat nodal ring approximately located at the Fermi level.



While the full Hamiltonian (\ref{eq:first-quant-H0}) will be the one used for analyzing Kohn's theorem in Appendix \ref{appendix:Kohn}, a linear approximation around the nodal line will be considered when computing the optical conductivity. This linear approximation is more easily written in the so-called toroidal coordinates $(k_r,\varphi_k,k_z)$. These are defined from the cylindrical coordinates $(K_r,\varphi_{k},K_z)$, but with the momenta measured from the nodal ring instead of the origin\footnote{Unless otherwise stated, we will use capital letters for momenta measured from the origin, while lower-case ones will be reserved for momenta measured from the nodal ring.}, i.e., $K_r = k_r + k_0$ and $K_z = k_z$. For $k_r,k_z \ll k_0$, we can expand the Hamiltonian up to linear order in momentum around the nodal ring, which amounts to retaining the leading order in a $1/k_0$ expansion\footnote{In fact, we will implicitly take the limit $k_0 \rightarrow \infty$ so that $k_r$ lies in the interval $(-\infty,\infty)$ instead of in $(-k_0,\infty)$.}, to obtain \cite{Rhim2016,Ahn2017}:
\begin{equation}
    \mathcal{H}_0(\boldsymbol{k}) = v_r k_r \tau_x + v_z k_z \tau_y,
    \label{eq:linear_Hamiltonian}
\end{equation}
where the radial Fermi velocity is $v_r = k_0 / m$ (for later convenience and with no physical effects, 
we have changed sign the term multiplying $\tau_x$ and interchanged $\tau_z$ by $\tau_y$). 

Furthermore, it is important to keep in mind that we will consider an isotropic linear dispersion in order to keep the analytical tractability of the problem. We therefore set $v_z=v_r=v_0$ in model (\ref{eq:linear_Hamiltonian}), leaving the study of the anisotropy for the future.

The linearized Hamiltonian (\ref{eq:linear_Hamiltonian}) allows the nodal line to be regarded as an infinite collection of 2D Dirac dispersions, each defined for a 2D $(k_r,k_z)$-plane in momentum space corresponding to a given azimuthal angle $\varphi_{k}$, which perpendicularly intersects the nodal ring. Loosely speaking, a NLSM can thus be thought of a large-$N$ graphene living in 3D. The similarity of this effective 2D dispersion of the NLSM with graphene will allow us to take advantage of the results obtained for this extensively studied material. Finally, let us point out that this linear approximation is valid as long as momentum and frequency are much smaller than $k_0$ and $v_r k_0$, respectively (see Fig. \ref{fig:Band_structure}). Consequently, in physical grounds, NLSMs with large nodal rings or open nodal lines (i.e., with large radius of curvature $k_0$) are ideal systems to observe the effects described in this work.




\section{\label{sec:optical_conductivity}Optical conductivity}

In Appendix \ref{appendix:Kohn} we provide a demonstration of the failure of Kohn's theorem in NLSMs. In this section, we examine one of the physical consequences of this failure: the non-zero interaction corrections to the conductivity. In particular, we have computed the uniform optical conductivity of the NLSM up to first order in the Coulomb interaction in the collisionless regime (i.e., with the frequency $\Omega$ being much larger than the disorder-induced scattering rate $1/\tau$, $\Omega \tau \gg 1$) at zero temperature and assuming the chemical potential $\mu$ to be pinned exactly at the nodal line. We are mainly interested in this intrinsic case, $\mu = 0$, since minimal screening will occur and thus a higher effect of Coulomb interaction is expected. Conversely, based on Fermi liquid's theory, whenever $\mu$ is placed sufficiently away from the line node, strong screening of the Coulomb interaction is expected to make it effectively short-ranged.

The diagrammatic calculations of the optical conductivity in terms of renormalized perturbation theory are presented in Appendix \ref{appendix:OC}. Here we will only summarize the results and provide their physical interpretation. 


As mentioned in the introduction, the optical conductivity is found to be:
\begin{subequations}
\begin{align}
     & \sigma_{zz}(\Omega) = \sigma_0 \left[1 + C_2 \alpha_R(\Omega) + \mathcal{O} \hspace{-2pt}\left(\alpha_R^2\right)\right] \hspace{-2pt}, \\
    & \sigma_{xx}(\Omega) = \sigma_{yy}(\Omega) = \frac{1}{2} \sigma_{zz}(\Omega).
\end{align}\label{eq:optical_conductivity_results}
\end{subequations}
\hspace{-3pt}Let us point out that $\sigma_{ij}$ will be multiplied by the degeneracy of the nodal loop (e.g. by 2 for a Dirac nodal line). We will first analyze the noninteracting part and then discuss the implications of the interaction correction.

As was already obtained by \cite{Ahn2017}, the noninteracting optical conductivity is $\sigma_0=k_0 e^2 / (16 \hbar)$. First of all, let us note that it is frequency independent (as long as the collisionless regime and our linear low-energy model apply, i.e., $1/\tau \ll \Omega \ll v_r k_0$), which is also characteristic of massless Dirac fermions in 2D. As proposed by \cite{Carbotte2017}, this feature can be used to distinguish the NLSM from Weyl and Dirac semimetals, where the optical conductivity grows linearly with frequency 
\cite{Roy2017}. Let us mention that for $\Omega \gg v_r k_0$ a linear dependence with frequency is expected by analogy with Weyl and Dirac semimetals (in our particular model, this linear increase applies only to  optical conductivity components parallel to the nodal line, with the perpendicular one tending to a constant value \cite{Barati2017}, by analogy with double-Weyl semimetals \cite{Lai2015,Li2016}). On the other hand, a Drude peak will arise in the low-frequency regime $\Omega \tau \ll 1$ \cite{Mukherjee2017,Barati2017}, where disorder, temperature and chemical potential have an important effect. 


Another interesting feature of the optical conductivity in the noninteracting limit is that it is determined by the product of a universal constant $e^2/(16 h)$, independent of material parameters, and the material-dependent nodal-ring perimeter $2 \pi k_0$. This contrasts with graphene, where the noninteracting optical conductivity, and thus the absorption, is universal \cite{Nair2008,Kuzmenko2008}. This difference is expected by dimensional analysis and analogy with graphene. In fact, the 3D optical conductivity, being a current \textit{density} response, has an additional inverse length unit compared to the corresponding 2D case. The only two quantities, within the low-energy effective model, with inverse length units in the uniform limit $\boldsymbol{q}\rightarrow0$ are $k_0$ and $\Omega/v$, the former being the one that appears in the response of a NLSM. Incidentally, only the quantity $\Omega/v$ is available for Weyl and Dirac semimetals, explaining the linear dependence of their conductivity with the frequency. The fact that it is $k_0$ and not $\Omega/v$ that enters in the NLSM response can be intuitively understood from noticing that the NLSM dispersion can be viewed as a collection of 2D Dirac dispersions, one for each point in the nodal line, so a proportionality to the nodal ring perimeter is expected.

Furthermore, the optical conductivity is highly anisotropic: the perpendicular component to the nodal ring is two times larger than the others. This result is in agreement with the optical response of a straight nodal line \cite{Shao2019}, where the components of the optical conductivity perpendicular to the nodal line are a non-zero constant while the parallel one vanishes. In our circular nodal line, the angular integration over the circumference gives rise to a factor $1/2$ in the parallel components. 

Let us now discuss the modifications induced by the Coulomb interaction. In the large-$k_0$ limit, the interaction correction turns out to be $\sigma_0 C_2 \alpha_R(\Omega)$, so, while being non-universal, it displays just the same material dependence as the non-interacting conductivity through the factor $k_0$ in $\sigma_0$. Moreover, it is characterized by exactly the same universal constant as in graphene, which we have determined to be $C_2 = (19-6 \pi)/12 \simeq 0.013$, the value more accepted up to date first obtained by \cite{Mishchenko2008}. Its smallness stems from the quasi-cancellation of the self-energy and vertex corrections to the polarization tensor. Let us point out that remaining at the leading order in the $1/k_0$-expansion is justified by the fact that our low-energy linear model is only valid for momenta and frequencies much smaller than $k_0$ and $v_r k_0$, respectively. This allows $k_0$ to be interpreted as the momentum UV-cutoff $\Lambda$, $\Lambda \sim k_0$, which must be very large for the renormalization method to work in our model. In other words, neglecting the next terms in $1/k_0$ is at the same level that neglecting the band bending through the quadratic dispersion.



The most remarkable effect of the interactions is the introduction of a logarithmic dependence on the frequency through the renormalized, i.e., physically measurable, coupling constant $\alpha_R(\Omega) = e^2/ [4 \pi \varepsilon \hbar v_R(\Omega)]$, with $v_R(\Omega)$ the renormalized Fermi velocity \cite{Gonzalez1994,Huh2016}. The logarithmic increase of $v_R(\Omega)$ with decreasing frequency implies that the coupling constant $\alpha_R(\Omega)$ displays a logarithmic decrease with decreasing frequency $\Omega$. Indeed, using the solution of the renormalization group equation (\ref{eq:beta_RG_v}), which relates $v_R$ at any frequency $\Omega$ to its value at other arbitrary energy $\bar{\mu}$ (the so-called renormalization scale) through $v_R(\Omega) = v_R(\bar{\mu}) [1 + (1/4) \alpha_R(\bar{\mu}) \ln{(\bar{\mu} / \Omega)} ]$, we find the following frequency dependence of $\alpha_R(\Omega)$:
\begin{equation}
    \alpha_R(\Omega) = \frac{\alpha_R(\bar{\mu})}{1 + (1/4) \alpha_R(\bar{\mu}) \ln{(\bar{\mu} / \Omega)}} \sim \frac{4}{\ln{(E_{\Lambda} / \Omega)}},
\end{equation}
where in the second approximation we have chosen the renormalization point to be the frequency UV-cutoff $E_{\Lambda}$ of the theory, and we have used that $\Omega \ll E_{\Lambda}$. 


\begin{figure}[!t]
    \centering
    \includegraphics[width=.35\textwidth]{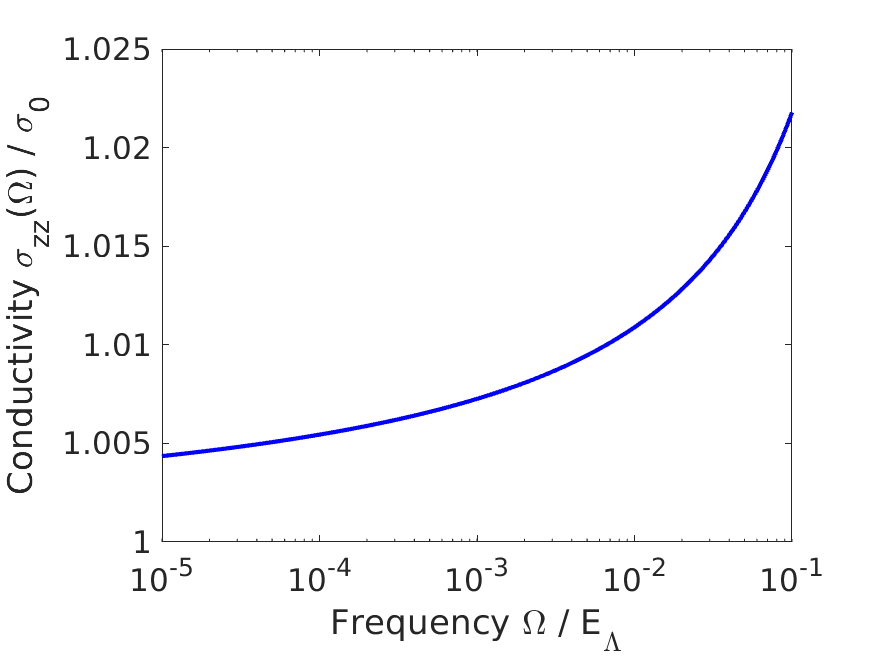}
    \caption{Representation of the optical conductivity $\sigma_{zz}$ (in units of $\sigma_0$) as a function of the frequency $\Omega$ (in units of the UV-cutoff $E_{\Lambda}$).}
    \label{fig:optical_conductivity}
\end{figure} 

Again, it is interesting to highlight the differences with 3D Weyl semimetals. In this case, the interaction correction is proportional to $\alpha_R(\Omega) [C_a + C_b \ln{(E_{\Lambda}/\Omega)}] \sim [C_a + C_b \ln{(E_{\Lambda}/\Omega)}]/\ln{(E_{\Lambda}/\Omega)} \sim C_b$, with $C_a$ and $C_b$ non-zero constants, i.e., the leading correction is approximately constant in frequency \cite{Rosenstein2013,Roy2017} (more precisely, this is the correction to the linear frequency dependence mentioned before). The appearance of this additional term $C_b \ln{(E_{\Lambda}/\Omega)}$ that eventually cancels the logarithmic contribution from the renormalized coupling constant $\alpha_R(\Omega)$ has been associated to the violation of hyperscaling in 3D quantum critical points (QCPs) \cite{Roy2017}, which is accompanied by logarithmic corrections to the thermodynamic potentials \cite{domb1976phase}. The QCP in NLSMs does not satisfies hyperscaling, but the absence of the additional logarithmic term indicates that $C_b=0$ for NLSMs, which means that violation of hyperscaling is not a sufficient condition for obtaining a nonzero $C_b$. 

Finally, let us point out that, even though we have considered zero temperature $T$ and vanishing chemical potential $\mu$, our results will be approximately valid in the collisionless region of the quantum-critical regime of the nodal-line fluid (in the vicinity of an electron-doped to hole-doped Fermi liquid transition). This corresponds to the limit $\mu \ll k_B T \ll \hbar \Omega \ll \hbar E_{\Lambda}$, where the interband contribution dominates compared to the intraband one.

\section{\label{section:Discussion}Discussion: experimental consequences}

Let us now discuss the observable effects in experiments. Recent infrared spectroscopy experiments have already determined the optical conductivity (indirectly from the reflectivity) in NLSMs, such as ZrSiS \cite{Schilling2017}, NbAs$_2$ \cite{Shao2019} and YbMnSb$_2$ \cite{Qiu2019}. While almost flat, i.e., frequency-independent, regions in the real part of the optical conductivity have been found, there are many features to discuss in the real systems. 


First of all, the three mentioned materials display a spin-orbit-induced gap, and therefore there is no physical nodal line. It is true though that the gap is small in the three cases, $\Delta \sim 10 \hspace{3pt} \text{meV}$, so that the electronic wave functions retain information about the parent nodal line that would exist in the absence of spin-orbit coupling. Therefore, a NLSM model for the fictitious nodal line would approximately apply for frequencies greater than twice the gap. Secondly, the (fictitious) nodal line does not have circular shape over the Brillouin zone in these materials. For instance, it forms a diamond-shaped network in ZrSiS \cite{Habe2018} and a curve that crosses the Brillouin zone in NbAs$_2$ \cite{Shao2019}. The main change compared to the circular nodal ring is that each component of the optical conductivity is expected to be proportional to the (appropriately projected) length of the line node in the corresponding direction. Moreover, as mentioned in the Introduction, since fine-tuning would be needed to ensure that the nodal line has constant energy, the (fictitious) nodal line in the three mentioned materials is energy dispersive. As argued by \cite{Ahn2017,Shao2019}, this case could be approximately tackled by substituting the total length of the nodal line (its perimeter $2 \pi k_0$ in our case) by the effective length that is allowed to be proved at frequency $\Omega$ by Pauli blocking. This effective length would grow with frequency, introducing an increase of the optical conductivity, until reaching the total length at high enough frequency, above which the optical conductivity would approximately take on the constant value determined without considering the energy dispersion.




With all this in mind, the noninteracting model would still predict a frequency-independent conductivity for high enough frequencies, in particular, greater than the decay rate $1/\tau$, the gap $\Delta$, and the chemical potential $\mu$, as well as allowing to prove the whole length of the nodal line. Indeed, an almost flat region has been measured for the three materials. While for NbAs$_2$ and YbMnSb$_2$ the comparison with \textit{ab initio} calculations has shown that this frequency-independent region arises from the effectively 2D Dirac nodal-line dispersion, in the case of ZrSiS a detailed DFT+multi-orbital tight-binding study \cite{Habe2018} has revealed that the band structure is not enough to reproduce it, but it is the interplay with disorder in the sample that provides this flat character. However, it is not known how the Coulomb interaction might change this scenario where the full lattice dispersion relation is considered. We will leave this question for the future.

We now discuss the possible experimental consequences of the Coulomb interaction on the optical conductivity. Two facts have to be taken into account. Firstly, the interaction correction we have determined is quite small compared to the noninteracting value. Indeed, for typical Fermi velocities $v_R \sim 5 \cdot 10^5 - 10^6 \text{m/s}$ \cite{Singha2017} and static lattice dielectric constants $\varepsilon \sim 10-40$
, the coupling constant takes on values $\alpha_R \sim 0.05-0.5$. If we assume a perturbative scheme to be valid\footnote{Some NLSMs might display relatively strong correlation effects, e.g. ZrSiS \cite{Pezzini2018}, thus spoiling the applicability of a perturbative treatment in principle.}, then the ratio of the interaction correction to the noninteracting value is $C_2 \alpha_R \sim 0.0005 - 0.005 \ll 1$. The second aspect to consider is that material-specific characteristics, as discussed in previous paragraphs, give important contributions to the optical conductivity, which our simple toy model does not capture.


What is otherwise expected, irrespective of some material-specific features, is the logarithmic dependence with frequency. After all, this essentially comes from the effective 2D Dirac dispersion, which is a good approximation whenever the curvature of the nodal line in the Brillouin zone is not large. Nevertheless, the change of the interaction correction over frequency is also quite small. For example, when the frequency is changed from $\Omega \sim 0.001 E_{\Lambda}$ to $\Omega \sim 0.1 E_{\Lambda}$, the conductivity increases about $\sim 1\%$ (see Fig. \ref{fig:optical_conductivity}). 
In any case, due to the complexity of features significantly contributing to the optical conductivity as well as to the relatively small frequency range ($\sim 10-100 \hspace{1pt} \text{meV}$) in which it is nearly flat (compared with $\sim 1 \hspace{1pt} \text{eV}$ in graphene \cite{Nair2008,Kuzmenko2008,Sheehy2009}), such expected logarithmic dependence is hidden in current experiments. Probably, if a real material better described by a simple model is found, the interaction correction could be measured, provided that experimental precision is high enough. In this respect, Ca$_3$P$_2$ is a good candidate \cite{Xie2015,Chan2016}.



\section{\label{section:conclusions}Conclusions}

To summarize, we have shown that electron-electron interactions induce corrections to the conductivity in NLSMs. 
By applying the field-theoretic perturbative renormalization procedure, we have determined the effect of the long-range Coulomb interaction on the optical conductivity. Our result applies to the frequency range where one can neglect the disorder-driven relaxation (collisionless regime), the probable energy dispersion of the nodal line, its possible (spin-orbit induced) gap, Pauli blocking and the effects of a finite temperature, while still being able to approximate the dispersion as linear around the line node. 

A remarkable conclusion of our work is the fact that, despite the different dimensionality, some analogies between 3D NLSMs and 2D Dirac systems appear when interactions are considered. Indeed, provided that the curvature of the nodal line is sufficiently small and in the region where its dispersion is approximately linear, the interaction correction logarithmically increases with frequency and the first-order coefficient is universal and equals exactly that of graphene. In fact, by mapping the calculation of the correction in the NLSM to the corresponding problem in graphene and by applying dimensional regularization along the same lines as \cite{Teber2018}, the first-order coefficient has the value $C_2 = (19-6 \pi)/12 \simeq 0.013$ of \cite{Mishchenko2008}. Our result also shows a fundamental interest due to enabling to differentiate the interaction effect in distinct 3D topological semimetals, such as Weyl semimetals and NLSMs.



Finally, regarding the experimental consequences of our work, even if our result is compatible with the experimental uncertainties, there is no evidence for a logarithmic frequency dependence due to the complexity of the band structure of the already known materials exhibiting nodal lines. Consequently, more work is needed in this line, both experimental, trying to find new materials behaving as simpler NLSMs (or trying to simulate transport experiments with cold atoms or photonic metamaterials), and theoretical, making predictions for more realistic models (e.g. including Fermi velocity anisotropy, energy dispersion of the nodal line, band-bending, chemical potential, temperature, etc.).  

\section*{\label{sec:Acknowledgement}Acknowledgements}

The authors are especially grateful to M. A. H. Vozmediano for invaluable comments and discussions.
D.M.S. was supported by CSIC JAE-Intro Grant No. JAEINT 18 01030 and by the Basque Government's grant PIBA 19-0081. A.C. acknowledges financial support through MINECO/AEI/FEDER, UE Grant No. FIS2015-73454-JIN and European Union structural funds, the Comunidad Aut\'onoma de Madrid (CAM) NMAT2D-CM Program (S2018-NMT-4511), and the Ram\'on y Cajal program through the grant RYC2018-023938-I.




\appendix

\section{\label{appendix:Kohn}Kohn's theorem}

This appendix is devoted to the analysis of the fate of Kohn's theorem in NLSMs. Kohn's theorem \cite{Kohn1961} is a powerful result imposing restrictions on the effect that electron-electron interactions might have on the long-wavelength conductivity and the cyclotron resonance frequency (if a magnetic field is applied). It states that in a Galileo-invariant system, i.e., a single-band Fermi-liquid metal with parabolic band dispersion and strictly obeying translational invariance, these two physical observables cannot be changed by interactions. Consequently, the conductivity in materials verifying Kohn's theorem may only be changed by processes explicitly breaking the translational symmetry or implying several bands \cite{Throckmorton2018}, such as Umklapp scattering due to the lattice \cite{Ashcroft1976}, Baber scattering associated with multiband systems \cite{Baber1936}, electron-hole scattering \cite{Kane1992}, electronic screening of impurities \cite{DasSarma1999}
, and Altshuler-Aronov-type interaction corrections in the presence of disorder \cite{Altshuler1983}. 
Indeed, the non-renormalization of the conductivity by interactions in Galileo-invariant Fermi-liquids is well known (technically, the self-energy and vertex corrections cancel each other) \cite{bruus2004many,Rosch2005,Maslov2012}.

The intuition behind the non-renormalization of the conductivity can be explained as follows. In a Galileo-invariant system, the total velocity, and thus the current, is proportional to the total momentum, with total referring to the sum for all the electrons. Since electron-electron interactions conserve the total momentum, they cannot alter the current. In materials where the total velocity is no longer proportional to the total momentum, momentum conservation does not imply current conservation. For instance, linearly-dispersing Weyl and Dirac semimetals have been shown to violate Kohn's theorem \cite{Throckmorton2018}, and therefore interactions affect their conductivity intrinsically without needing explicit breaking of the translational symmetry. 

For the NLSM, a na\"{i}ve look at the Hamiltonian (\ref{eq:first-quant-H0}) might lead to the wrong conclusion that, due to the parabolic dispersion in the $x$ and $y$ directions, Kohn's theorem might partially apply in these directions or when the external magnetic field points in the $z$ direction. However, below we explicitly show that Kohn's theorem fails, leading to a non-vanishing correction of the conductivity by the electron-electron interactions. Aside from the deviation from an isotropic quadratic dispersion, we have identified the main cause of Kohn's theorem failure to be the presence of more than one band. We therefore argue that Kohn's theorem is in fact a very specific result and will in general not apply. It may though be a good approximation for isotropic Fermi-liquid metals in which the chemical potential is well inside one band and far away from the rest (compared to the rest of relevant energy scales).

Let us now present the rigorous proof of our previous assertion. For that, we will use the first-quantized version of the full Hamiltonian (\ref{eq:first-quant-H0}). Assume that an external magnetic field $\boldsymbol{B}$ is applied. Due to the initial rotational symmetry in the $xy$-plane in the absence of $\boldsymbol{B}$, we can choose the $x$ axis such that the most general magnetic field lies in the $xz$-plane, making an angle $\theta$ with the $z$ axis: $\boldsymbol{B} = B \sin(\theta) \boldsymbol{e}_x + B \cos(\theta) \boldsymbol{e}_z$. Without losing generality, we take the angle $\theta$ to range from 0 to $\pi/2$, with the interval $(\pi/2,\pi]$ considered by a negative $B$. In an appropriate Landau gauge, the vector potential lies in the $y$ axis: $\boldsymbol{A} = B \left[ x \cos(\theta) - z \sin(\theta) \right] \boldsymbol{e}_y$. The $N$-particle Hamiltonian minimally coupled to the external magnetic field is\footnote{Throughout the analysis of Kohn's theorem, Latin subindices will label the electron on which the operator acts, while Greek subindices will be reserved for spatial coordinates $x$, $y$, or $z$.}: 
\begin{equation}
\begin{split}
    H = & \sum_{i=1}^{N} \left[ \left( \Delta - \frac{P_{i,x}^2+P_{i,y}^2}{2 m} \right) \tau_{i,x} + v P_{i,z} \tau_{i,z} \right] + \\
    & + \sum_{1 \leq i < k \leq N} u(\boldsymbol{r}_i - \boldsymbol{r}_k),
\end{split}
\label{eq:many-body-H}
\end{equation}
where $u(\boldsymbol{r}_i - \boldsymbol{r}_k)$ is a two-body interaction between electrons $i$ and $k$ dependent on their relative position, and $\boldsymbol{P}_i = \boldsymbol{K}_i + e \boldsymbol{A}_i = \boldsymbol{K}_i + e B \left[ x_i \cos(\theta) - z_i \sin(\theta) \right] \boldsymbol{e}_y$ is the mechanical momentum of electron $i$. We assume that the two-body interaction is even, i.e., $u(\boldsymbol{r}) = u(- \boldsymbol{r})$. For instance, the Coulomb interaction $u(\boldsymbol{r}) = e^2/(4 \pi \varepsilon |\boldsymbol{r}|)$ verifies that, although our results apply to more general interactions. For shortness, we will sometimes use cylindrical coordinates such that $P_{i,r}^2=P_{i,x}^2+P_{i,y}^2$ as well as the simplified notation $u_{ik} \equiv u(\boldsymbol{r}_i - \boldsymbol{r}_k)$.

The steps followed in the subsequent analysis are the following. First, we calculate the velocity operator of one electron, from which the total velocity operator can be determined. The next step is writing the equations of motion (in the Heisenberg picture) for the one-particle and total mechanical momentum operators, which can be easily expressed in terms of the velocity operators. Finally, the equation of motion for the total velocity operator is written. If, as in the NLSM, in this last equation there exists a non-vanishing term explicitly containing the interaction potential, then Kohn's theorem is not fulfilled since interactions modify the total velocity and thus the total current, as well as the cyclotron frequency.

The velocity operator $\boldsymbol{v}_j = \frac{d \boldsymbol{r}_j}{dt} = \frac{i}{\hbar} [H,\boldsymbol{r}_j]$ of electron $j$ in Cartesian components is given by:
\begin{equation}
v_{j,x} = - \frac{P_{j,x}}{m}  \tau_{j,x} \hspace{3pt} , \hspace{3pt} v_{j,y} = - \frac{P_{j,y}}{m} \tau_{j,x} \hspace{3pt} , \hspace{3pt} v_{j,z} = v \tau_{j,z}.
\label{eq:velocityj-components}
\end{equation}
This is in fact the expected result according to the quadratic and linear dispersions. The presence of the Pauli matrices prevents the total velocity $\boldsymbol{v} = \sum_{j=1}^N \boldsymbol{v}_j$ from being proportional to the total mechanical momentum even in the $x$ and $y$ axes, which is the main reason why Kohn's theorem does not hold in our system.


To obtain the equation of motion for the mechanical momentum $P_{j,\beta}$, one needs the commutators:
\begin{align}
    & [P_{i,\alpha},P_{j,\beta}] = i \hbar e \delta_{ij} \left( \frac{\partial A_{j,\alpha}}{\partial r_{j,\beta}} - \frac{\partial A_{j,\beta}}{\partial r_{j,\alpha}} \right), \\
    & [u_{ik},P_{j,\beta}] = i \hbar \left\{ \delta_{ij} \partial_{\beta} u_{jk} + \delta_{kj} \partial_{\beta} u_{ji} \right\},
\end{align}
where we have used the results for the commutation relations for functions of operators obtained by \cite{Transtrum2005}, as well as the fact that $u(\boldsymbol{r})=u(-\boldsymbol{r})$ implies that $\partial_{\beta} u(\boldsymbol{r}) = -\partial_{\beta} u(-\boldsymbol{r})$, where $\partial_{\beta} u(\boldsymbol{r}) \equiv \frac{\partial u(\boldsymbol{r}')}{\partial r'_{\beta}}\big|_{\boldsymbol{r}}$. The equation of motion may be most suitably written in the cartesian coordinates $(\parallel,y,\perp)$ in which $\parallel$ and $y$ lie in the plane perpendicular to the external magnetic field and $\perp$ points in the direction of the magnetic field. For any vector $\boldsymbol{W}$, 
\begin{equation}
\begin{split}
    & W_{\perp} = \cos(\theta) W_z + \sin(\theta) W_x, \\
    & W_{\parallel} = - \sin(\theta) W_z + \cos(\theta) W_x.
\end{split}
\end{equation}
We furthermore define $W_{\pm} = W_{\parallel} \pm i W_y$, which are (proportional to) the raising and lowering operators for the Landau levels. The equation of motion for the one-particle mechanical momentum operator then reads as:
\begin{equation}
\begin{split}
    & \frac{d P_{j,\pm}}{d t} = \pm i e B v_{j,\pm} - \sum_{\substack{k=1 \\ k \neq j}}^{N} 2 \partial_{\mp} u_{jk}, \\[-10pt]
    & \frac{d P_{j,\perp}}{d t} = - \sum_{\substack{k=1 \\ k \neq j}}^{N} \partial_{\perp} u_{jk}, 
\end{split}
\label{eq:momentajpm}
\end{equation}
where $\partial_{\mp} u(\boldsymbol{r}) \equiv \frac{\partial u(\boldsymbol{r}')}{\partial r'_{\mp}}\big|_{\boldsymbol{r}}$ and $\partial_{\perp} u(\boldsymbol{r}) \equiv \frac{\partial u(\boldsymbol{r}')}{\partial r'_{\perp}}\big|_{\boldsymbol{r}}$, with $r_{\mp} = r_{\parallel} \mp i y$. Summing up for all the electrons, the total mechanical momentum verifies:
\begin{equation}
\frac{d P_{\pm}}{d t} = \pm i eB v_{\pm} \hspace{10pt} , \hspace{15pt} \frac{d P_{\perp}}{d t} = 0.
\label{eq:totalPcomponents}
\end{equation}
Note that no many-body interactions appear explicitly due to the fact that $\sum_{j=1}^{N}\sum_{k=1;k\neq j}^{N} \partial_{\beta} u_{jk} = 0$ since $\partial_{\beta} u$ is odd with respect to the inversion operation $\boldsymbol{r} \rightarrow - \boldsymbol{r}$. 


Let us interpret Eq. (\ref{eq:totalPcomponents}). As usual, the magnetic field affects only the dynamics in the plane perpendicular to it, and the total momentum in the direction of the magnetic field $P_{\perp}$ is conserved. Furthermore, if we set $\boldsymbol{B}=0$, we recover the conservation of the total momentum. On the other hand, in the presence of a magnetic field, the cyclotron resonance frequency $\omega_c$ is defined by the expression $d P_{\pm}/d t = \pm i \omega_c P_{\pm}$. In this case, due to the lack of proportionality between momenta and velocity, we cannot conclude that the total momentum and the total velocity would be unaffected by the electron-electron interactions. In fact, many-body interactions change the evolution of the total velocity, which implies a renormalization of the current and, from its definition together with Eq. (\ref{eq:totalPcomponents}), also of the cyclotron resonance frequency.

In order to see explicitly the presence of the many-body interactions in the equations of motion, we need to calculate the second derivative of the momentum, which is proportional to the equation of motion for the total velocity $v_{\pm}$. With a view to avoiding mathematical difficulty and focusing on the physical interpretation, let us first compute this for the two particular cases in which the magnetic field points in the $z$ and $x$ directions, and finally state the results for the general case.

\subsection{Magnetic field in the $z$ direction}

In this case, the angle of the magnetic field with the $z$ axis is $\theta=0$, so that $v_{\parallel} = v_x$, $v_{\perp} = v_z = v \sum_{j=1}^N \tau_{j,z}$ and $v_{\pm} = - \sum_{j=1}^N (P_{j,\pm}/m) \tau_{j,x}$. Using expressions (\ref{eq:momentajpm}) and the following time derivatives of the Pauli matrices:
\begin{align}
    & \frac{d \tau_{j,x}}{dt} = -2 \frac{1}{\hbar} v P_{j,z} \tau_{j,y} = - 2 \frac{i}{\hbar} v_{j,z} P_{j,z} \tau_{j,x} \tau_{j,z}, \\
    & \frac{d \tau_{j,z}}{dt} = i \frac{2}{\hbar} \left( \Delta - \frac{P_{j,x}^2+P_{j,y}^2}{2m} \right) \tau_{j,x} \tau_{j,z},
\end{align}
we find the equations of motion for the total velocity:
\begin{align}
    \begin{split}
    \frac{d v_{\pm}}{dt} & = \pm i \frac{e B}{m} \frac{P_{\pm}}{m} + i \frac{2}{\hbar} \sum_{j=1}^N v_{j,\perp} v_{j,\pm} P_{j,\perp} + \\[-3pt]
    & \hspace{12pt} + \frac{2}{m} \sum_{j=1}^N \sum_{\substack{k=1 \\ k \neq j}}^N \partial_{\mp} u_{jk} \tau_{j,x}, 
\end{split} \label{eq:dvpmBz} \\ 
    \frac{d v_{\perp}}{dt} & = - i \frac{2}{\hbar} \sum_{j=1}^N v_{j,\perp} \left( \Delta - \frac{P_{j,r}^2}{2m} \right) \tau_{j,x}. \label{eq:dvperpBz} 
\end{align}
The presence of the Pauli matrix $\tau_{j,x}$ multiplying the derivatives of the many-body interaction in the $v_{\pm}$ equation (\ref{eq:dvpmBz}) implies that this sum does not vanish. In fact,
\begin{equation}
\begin{split}
    2 \sum_{j=1}^N \sum_{\substack{k=1 \\ k \neq j}}^N & \partial_{\mp} u_{jk} \tau_{j,x} = \\[-10pt]
    & \hspace{-5pt} = \frac{1}{2} \sum_{j=1}^N \sum_{\substack{k=1 \\ k \neq j}}^N (\tau_{j,x}-\tau_{k,x}) [\partial_{x} u_{jk} + \partial_{y} u_{jk}], \hspace{-2pt}
\end{split}
\end{equation}
which is distinct from zero in general since the terms being summed up are even under the exchange of indices $j \leftrightarrow k$. On the contrary, interactions do not enter explicitly in the $v_{\perp}$ equation (\ref{eq:dvperpBz}). However, the second derivative $d^2 v_{\perp}/dt^2$ contains terms proportional to $\sum_{j=1}^N \sum_{k=1 ; k \neq j}^N P_{j,\alpha} \partial_{\alpha} u_{jk} \tau_{j,y}$, with $\alpha=x,y$, which again do not vanish. 
Indeed,
\begin{equation}
\begin{split}
    \frac{d^2 v_{\perp}}{d t^2} = & \hspace{2pt} i \left( \frac{2}{\hbar} \right)^2 v \sum_{j=1}^N \left( \Delta - \frac{P_{j,r}^2}{2m} \right) H_j^0 \tau_{j,y} + \\[-3pt]
    & + \frac{2 v}{\hbar m} \sum_{j=1}^N \sum_{\substack{k=1 \\ k \neq j}}^N [P_{j,x} \partial_x u_{jk} + (x \leftrightarrow y)] \tau_{j,y}, 
\end{split}
\end{equation}
where $H_j^0 = \left( \Delta - \frac{P_{j,r}^2}{2 m} \right) \tau_{j,x} + v P_{j,z} \tau_{j,z}$ is the non-interacting part of the Hamiltonian of the $j$ electron.



Therefore, we have shown that, when the magnetic field points parallel to the $z$ axis, the evolution of the total velocity depends on the many-body interactions, and consequently both the current and the cyclotron resonance frequency will be renormalized by electron-electron interactions in NLSMs. Let us point out that this result comes mathematically from the presence of the Pauli matrices, which have a non-commutative algebra. Physically, this means that it is the presence of the two bands that gives rise to the violation of the Kohn's theorem. We anticipate that the same result will be obtained for an arbitrary magnetic field.


\subsection{Magnetic field in the $x$ direction}

In this case, the angle of the magnetic field with the z axis is $\theta=\pi/2$, so that $v_{\parallel} = - v_z$, $v_{\perp} = v_x = - \sum_{j=1}^N \frac{P_{j,x}}{m} \tau_{j,x}$, and $v_{\pm} = - \sum_{j=1}^N \left( v \tau_{j,z} \pm i \frac{P_{j,y}}{m} \tau_{j,x} \right)$. Following analogous steps as in the previous section, we find that the equation of motion for the total velocity is: 
\begin{align}
\begin{split}
    & \frac{d v_{\pm}}{dt} = \sum_{j=1}^N \frac{v_{j,z}}{\hbar} \left[ -2i \left( \Delta - \frac{P_{j,r}^2}{2m} \right) \tau_{j,x} \pm \right. \\
    & \hspace{26pt} \left. \vphantom{\frac{P_{j,r}^2}{2m}} \pm \{ v_{j,y} , P_{j,z} \} \right] \pm \frac{i}{m} \sum_{j=1}^N \sum_{\substack{k=1 \\ k \neq j}}^N \partial_{y} u_{jk} \tau_{j,x}, 
\end{split} \\
\begin{split}
    & \frac{d v_{\perp}}{dt} = i \frac{2}{\hbar} \sum_{j=1}^N v_{j,z} v_{j,x} P_{j,z} + \frac{1}{m} \sum_{j=1}^N \sum_{\substack{k=1 \\ k \neq j}}^N \partial_x u_{jk} \tau_{j,x}. 
\end{split} \label{eq:vperpBx}
\end{align}
Again, due to the presence of the Pauli matrices, the terms containing explicitly the many-body interactions do not vanish. However, contrary to the case in which $\boldsymbol{B} = B \boldsymbol{e}_z$, the non-vanishing term due to the interactions in the $v_{\perp}$ equation (\ref{eq:vperpBx}) already appears at the first time-derivative. This asymmetry stems from the different dispersion relation in the $x$ and $z$ directions. In fact, Ref. \cite{Throckmorton2018} showed that the first non-vanishing explicit interaction-dependent term already appeared at the first time-derivative of the velocity operator in the case of bilayer graphene (quadratic band touching), while one should calculate the second time-derivative of the velocity when dealing with Weyl semimetals (linear dispersion) to see this term appear.

\subsection{Arbitrary magnetic field}

Let us now calculate the time derivative of the velocity operators in the general case when $\boldsymbol{B} = B [\cos(\theta) \boldsymbol{e}_z + \sin(\theta) \boldsymbol{e}_x]$. Given the particular results discussed above, we expect the Kohn's theorem to fail also in this general case, since a non-vanishing interaction-dependent term is expected to appear at the first time-derivative of the velocity operators, as we will now show. The only increased difficulty of this general case compared to the previous particular cases arises from the lengthier mathematical expressions. Given that expressing the results in terms of just the $\perp$ and $\pm$ components of the velocities and momenta is more complex and it does not provide an easier interpretation, we will provide the results in terms of both the $\perp$, $\pm$ and the $x,y,z$ components. 

After following the same steps as in the previous cases, we arrive at the following expressions for the time-derivatives of the total velocity operators:
\begin{align}
\begin{split}
\frac{d v_{\perp}}{dt} & = \sum_{j=1}^N \left\{ \frac{2i}{\hbar} v \left[ \cos(\theta) \left( \Delta - \frac{P_{j,r}^2}{2m}\right) \tau_{j,x} - \right. \right. \\[-2pt]
& \hspace{-10pt} - \hspace{-4pt} \left. \left. \vphantom{\frac{P_{j,r}^2}{2m}} \sin(\theta) P_{j,z} v_{j,x} \right] \tau_{j,z} \right\} \hspace{-1pt} + \sin(\theta) \cos(\theta) \frac{eB}{m} \frac{P_{y}}{m} + \\[-2pt]
& \hspace{-10pt} + \hspace{-1pt} \frac{1}{m} \sin(\theta) \sum_{j=1}^N \sum_{\substack{k=1 \\ k \neq j}}^N \partial_x u_{jk} \tau_{j,x}, 
\end{split} 
\label{eq:general-dvperpdt}
\end{align}
and:
\begin{align}
\begin{split}
\frac{d v_{\pm}}{dt} & = \frac{eB}{m} \left[ - \sin^2(\theta) \frac{P_{y}}{m} \pm i \sum_{j=1}^N v_{j,\pm} \tau_{j,x} \right] - \\[-2pt]
& \hspace{3pt} - \frac{2i}{\hbar} v \sum_{j=1}^N \left\{ \left[ \left( \cos(\theta) v_{j,x} \pm i v_{j,y} \right) P_{j,z} + \vphantom{\frac{P_{j,r}^2}{2m}} \right. \right. \\[-2pt]
& \hspace{3pt} + \left. \left. \sin(\theta) \left( \Delta - \frac{P_{j,r}^2}{2m} \right) \tau_{j,x} \right] \tau_{j,z} \right\} + \\[-2pt]
& \hspace{3pt} + \hspace{-1pt} \frac{\sin(\theta)}{m} \hspace{-1pt} \sum_{j=1}^N \sum_{\substack{k=1 \\ k \neq j}}^N \hspace{-1pt} \left[ \cos(\theta) \partial_x u_{jk} \pm i \partial_y u_{jk} \right] \tau_{j,x}. 
\end{split}
\label{eq:general-dvpmdt}
\end{align}
Analogously to the previous cases, the terms explicitly featuring the electron-electron interactions do not vanish due to the algebra of the Pauli matrices, i.e., the presence of two bands in which the interband transitions have to be taken into account. Incidentally, it is straightforward to verify that these general expressions reduce to the particular cases calculated above. Note that the only value of $\theta$ for which the interaction-dependent term vanishes in Eq. (\ref{eq:general-dvperpdt}) is $\theta = 0$, i.e., when the magnetic field points towards the $z$ direction, where we proved that the interactions enter explicitly into the second time-derivative.


As we had previously advanced, we have explicitly proved that Kohn's theorem fails in NLSMs. In the absence of magnetic field, the conductivity is renormalized by electron-electron interactions in such systems even if the total momentum is conserved. When a magnetic field is introduced, both the dynamics of the total momentum and total velocity are changed by these many-body interactions, which results in a renormalization of the cyclotron resonance frequency too.

The key point for the verification of Kohn's theorem can be deduced from Eq. (\ref{eq:totalPcomponents}), which can be shown to hold for a general Hamiltonian. Accordingly, Kohn's theorem will hold provided that the total velocity is a conserved quantity. For example, it is interesting to note that even for the simplest quadratic two-band Hamiltonian $H_0 = \left[ \boldsymbol{P}^2/(2 m) - \Delta \right] \tau_z$ Kohn's theorem fails (electron-hole scattering prevents its verification).

\section{ \label{appendix:OC}Optical conductivity}

This appendix presents the perturbative calculations of the uniform optical conductivity of the isotropic-Fermi-velocity NLSM up to first order in the Coulomb interaction in the collisionless regime at zero temperature and vanishing chemical potential. For this one needs the polarization tensor or photon self-energy $\Pi_{\mu \nu}$, where\footnote{We use the convention of e.g. \cite{schakel2008boulevard,Teber2014,Teber2018} to define the polarization tensor, which differs in a minus sign from e.g. \cite{bruus2004many,Rhim2016}.} $\Pi_{00}(\Omega,\boldsymbol{q}) = i e^2 \langle \rho(\Omega,\boldsymbol{q}) \rho(-\Omega,-\boldsymbol{q}) \rangle$ is the charge density-density response function and $\Pi_{ij}(\Omega,\boldsymbol{q}) = i e^2 \langle j_i(\Omega,\boldsymbol{q}) j_j(-\Omega,-\boldsymbol{q}) \rangle$ is the charge current-current response function\footnote{Since we implicitly work with the functional integral formalism, the expectation value \unexpanded{$ \langle \cdot \cdot \cdot \rangle $} is implicitly indicating time ordering.}, with $\Omega$ the frequency, $\boldsymbol{q}$ the wave vector and $e$ the electron charge. 
It is well known from linear response theory that the conductivity tensor $\sigma_{ij}$ and the polarization are related through $\sigma_{ij}^R(\Omega,\boldsymbol{q}) = -i (1/\Omega) \Pi_{ij}^R(\Omega,\boldsymbol{q})$ (the superscript $R$ indicates that we are interested in the retarded response due to causality).

To obtain the particle current density operator $\hat{j}_{i}$, $i=x,y,z$, we minimally couple the linear Hamiltonian (\ref{eq:linear_Hamiltonian}) to an \textit{external classical} electromagnetic potential $A_{i}^{ext}$ by substituting $k_i \rightarrow k_i + e A_{i}^{ext}$. Then, $\hat{j}_{i}$ can be extracted from the functional derivative \cite{bruus2004many}:%
\small%
\setlength{\abovedisplayskip}{10pt}%
\setlength{\belowdisplayskip}{10pt}%
\begin{equation}
     \hat{j}_i (x) \hspace{-2pt}  = \hspace{-2pt} \frac{1}{e} \frac{\delta \hat{H}_0}{\delta A_i^{ext}(x)} \hspace{-2pt} = \hspace{-1pt} \hat{a}^{\dagger}(x) j_i \hat{a}(x) \hspace{1pt} ; \begin{cases} \hspace{-1pt} j_x \hspace{-2pt} = \hspace{-1pt} v_r \hspace{-1pt} \cos(\varphi_{k}) \tau_x, \\ \hspace{-1pt} j_y \hspace{-2pt} = \hspace{-1pt} v_r \hspace{-1pt} \sin(\varphi_k) \tau_x, \\ \hspace{-1pt} j_z \hspace{-2pt} = \hspace{-1pt} v_z \tau_y, \end{cases} \hspace{-5pt}
    \label{eq:current_densities}
\end{equation}%
\normalsize%
where $v_r=v_z=v_0$ in our isotropic-Fermi-velocity model, $\cos(\varphi_{k}) = k_x/k_r$ and $\sin(\varphi_{k}) = k_y/k_r$ (note that they are operators if one works in the position representation). In the same way, we obtain the particle density $j_0 \equiv \rho = \tau_0$. Also notice that, within our linear approximation, there is no diamagnetic current, since the terms quadratic in $A_i^{ext}$ would appear at order $\mathcal{O}(1/k_0)$. 




We start by defining the action $\mathcal{S}$ for the fermionic quantum field $\psi_0(x)$ for the isotropic-Fermi-velocity NLSM. We will consider the chemical potential to be pinned exactly at the nodal line, i.e., $\mu=0$. We also assume zero temperature, so that we can work with the real time formalism. 
The four-momenta are defined with the Fermi velocity in the spatial components, $k=(\omega, v_0 \boldsymbol{k})$. By means of a Hubbard-Stratonovich transformation \cite{coleman2015introduction}, the quartic instantaneous 3D Coulomb interaction $\rho(q) U(|\boldsymbol{q}|) \rho(-q)$, with $U(|\boldsymbol{q}|) = e_0^2/(\varepsilon |\boldsymbol{q}|^2) \equiv g_0^2/(|\boldsymbol{q}|^2)$ and $g_0 = e_0/\sqrt{\varepsilon}$ the effective charge in the NLSM, can be substituted by a coupling of the fermionic spinor field with a photonic scalar field $A_0$ of the form $\psi^\dagger_0(x) g_0 A_0(x) \psi_0(x)$. After performing this transformation, the action, in units such that $\hbar = c = 1$, reads as $\mathcal{S} = \mathcal{S}_0 + \mathcal{S}_A + \mathcal{S}_{int}$, where:
\begin{align}
    & \mathcal{S}_0 = \int d^4 x \psi^\dagger_0(x) \left[ \tau_0 \omega - v_0 \boldsymbol{\tau} \cdot \boldsymbol{k} \right] \psi_0(x), \\
    & \mathcal{S}_A = \int d^4 x \frac{1}{2}  [\boldsymbol{\nabla}_{3D} A_0(x) ]^2, \\
    & \mathcal{S}_{int} = \int d^4 x \psi^\dagger_0(x) \tau_0 (-g_0) A_0(x) \psi_0(x),
\end{align}
with $\tau_0$ the $2 \times 2$ identity matrix, $\boldsymbol{\tau}=(\tau_x,\tau_y)$, $\omega = i\partial_t$,  $\boldsymbol{k}=(k_r,k_z)=(-i\partial_r,-i \partial_z)$, and $\boldsymbol{\nabla}_{3\text{D}}=(\partial_x,\partial_y,\partial_z)$. All the parameters appearing in this action, i.e., $g_0$, $v_0$, $\psi_0$ and $A_0$, are bare, unrenormalized ones. 



In this work, we will apply renormalized perturbation theory via dimensional regularization (DR) \cite{veltman1972regularization,schwartz2014quantum,collins1985renormalization}. The motivation for using renormalized perturbation theory is the following. If bare, normal perturbation theory in the Coulomb interaction were to be carried out, some results would turn out to be infinite (technically, some loops would diverge when integrating over large momenta). Physical results are finite, so these divergences must be an artifact of the calculation procedure. Indeed, they arise due to using a low-energy effective field theory, which is only valid up to some UV-cutoff, instead of the complete field theory (a lattice model, in our case). Renormalized perturbation theory \textit{properly} avoids these infinities to obtain the correct low-energy results. 

There are several regularization procedures to do so. Probably, the most intuitive one is introducing a hard UV-cutoff $\Lambda$ in the momentum integrations, which prevents the divergences. However, this simple method has some disadvantages, the most notable one being that it does not automatically guarantee gauge invariance. This is the main reason behind using DR. This procedure consists of extending the initial four-dimensional space-time to $\tilde{d} = 4 - 2 \epsilon$ dimensions, but keeping $\text{Tr}[\tau_0]=2$. In our case, we will see that it will be necessary to extend only the ($k_r,k_z$) subspace to a ($D=2-2\epsilon$)-dimensional space so as to get finite results, since the integrals in frequency and the perpendicular momentum will be finite. At the end of the calculations, one should take the $\epsilon \rightarrow 0$ limit. In DR, some diagrams will initially present divergent parts proportional to negative powers of $\epsilon$, which have to be properly substracted to ensure the finiteness of the correct physical results. 

In order to do so, we propose that the bare parameters appearing in the action are related to the physical, renormalized ones through $x_0 = Z_x x_R$, $x \equiv v,\psi,A$, where $Z_x$ are the so-called renormalization constants. It is conventional also to define the so-called counterterms $\delta_x$ from the renormalization constants $Z_x$ via $Z_x = 1 + \delta_x$. The previously mentioned divergences are absorbed into the appropriate counterterms, i.e., some bare parameters must be infinite so that the renormalized parameters and the physical results are finite. An important point is that the results must be properly renormalized order by order in perturbation theory. The counterterms are therefore defined as a perturbative series in the coupling constant (the electric charge in our case), their leading order being $\mathcal{O}(e_R^2)$ (in the noninteracting limit there are no divergences, so the bare parameters are equal to the renormalized ones, and thus $Z_x=1$). In summary, DR consists of considering the interactions perturbatively in $\tilde{d} = 4 - 2 \epsilon$ dimensions, and choosing the appropriate counterterms to be infinite in order to cancel the divergences. 

Regarding the electric charge, let us note that, even though in the original four dimensions it is dimensionless (which determines its marginal character at tree level), when extending to $\tilde{d}$-dimensions it acquires a mass dimension of $[g_0]=[e_0]=\epsilon$. In order to keep the renormalized electric charge dimensionless as in the original 4-dimensions, we will use the modified minimal subtraction ($\bar{\text{MS}}$) scheme \cite{Teber2014,Teber2018,schwartz2014quantum}. This procedure amounts to introducing a quantity with energy units, the so-called renormalization scale $\bar{\mu}$, in the renormalization of the coupling: $e_0  = Z_e e_R \bar{\mu}^{\epsilon} (4 \pi)^{-\epsilon/2} e^{\gamma_E \epsilon /2}$, where $\gamma_E \simeq 0.577$ is the Euler-Mascheroni constant.

With this in mind, the Lagrangian density can be written in terms of the renormalized parameters as $\mathcal{L} = \mathcal{L}_0 + \mathcal{L}_A + \mathcal{L}_{int}$, where:
\begin{align}
\begin{split}
        \mathcal{L}_0 \hspace{3pt} & = \psi^\dagger_0 [ \tau_0 \omega - v_0 \boldsymbol{\tau} \cdot \boldsymbol{k} ] \psi_0 =  \\ 
        & = \psi^\dagger_R [ \tau_0 \omega - v_R \boldsymbol{\tau} \cdot \boldsymbol{k} ] \psi_R + \\
        & \hspace{10pt} + \psi^\dagger_R [ \delta_{\psi} \tau_0 \omega - (\delta_{\psi}+\delta_v) v_R \boldsymbol{\tau} \cdot \boldsymbol{k} ] \psi_R,
\end{split} \\[5pt]
\begin{split}
         \mathcal{L}_A \hspace{3pt} & = \frac{1}{2}  [\boldsymbol{\nabla}_{3\text{D}} A_0 ]^2 \hspace{-2pt} = \hspace{-1pt} \frac{1}{2}  [\boldsymbol{\nabla}_{3\text{D}} A_R ]^2 \hspace{-2pt} + \hspace{-1pt} \frac{\delta_A}{2}  [\boldsymbol{\nabla}_{3\text{D}} A_R ]^2 \hspace{-2pt},
\end{split} \\[5pt]
\begin{split}
        \mathcal{L}_{int} \hspace{-1pt} & = - g_0 \psi^\dagger_0 \tau_0 A_0 \psi_0 = \bar{\mu}^{\epsilon} (4 \pi)^{-\epsilon/2} e^{\gamma_E \epsilon /2} \cdot \\
        & \hspace{12pt} \cdot \left[ - g_R \psi^\dagger_R \tau_0 A_R \psi_R - \delta_{\text{Coul}} g_R \psi^\dagger_R \tau_0 A_R \psi_R \right],
\end{split}
\end{align}
where $\delta_{\text{Coul}} = \delta_e + \delta_{\psi} + \delta_A/2 + \mathcal{O}(e_R^4)$. The Feynman rules can be derived from this Lagrangian density, with the counterterms being represented by their corresponding diagrams. Let us anticipate that the charge will not be renormalized to lowest order, $Z_e=1+\mathcal{O}(e_R^4)$, so that we can write $e_0 \equiv e_R \bar{\mu}^{\epsilon} (4 \pi)^{-\epsilon/2} e^{\gamma_E \epsilon /2}$ for simplicity.

We have then the following free electron propagator: 
\begin{equation}
\begin{split}
    & \begin{gathered} \includegraphics[scale=0.2]{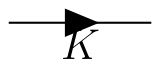} \end{gathered} = S_R^{(0)}(K) = \langle \psi_R(K) \psi^\dagger_R(K) \rangle =\\
    & = i (\tau_0 \omega - v_R \boldsymbol{\tau} \cdot \boldsymbol{k})^{-1} = i \frac{\tau_0 \omega + v_R \boldsymbol{\tau} \cdot \boldsymbol{k}}{\omega^2 - v_R^2 \boldsymbol{k}^2}.
\end{split}
\label{eq:free_electron_propagator}
\end{equation}
The free Coulomb photon propagator (which is instantaneous, so it does not depend on frequency) reads as:
\begin{equation}
    \begin{gathered} \includegraphics[scale=0.2]{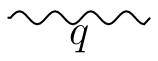} \end{gathered} = V_R^{(0)}(q) = \langle A_R(q) A^\dagger_R(q)\rangle = i \frac{1}{\boldsymbol{q}^2}.
    \label{eq:free_photon_propagator}
\end{equation}
The Coulomb interaction vertex, which only couples scalar photons, is:
\begin{equation}
    \begin{gathered} \includegraphics[scale=0.2]{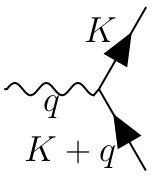} \end{gathered} \hspace{-2pt} = \hspace{-1pt} - i g_0 \tau_0 = \hspace{-1pt} -i g_0 \Gamma^{(0)}_{\text{Coul}} \hspace{2pt}, \text{ with} \hspace{2pt} \Gamma^{(0)}_{\text{Coul}} = \tau_0,
    \label{eq:Coulomb_interaction_vertex}
\end{equation}
while the coupling to the external field, which also couples vector photons, 
can be extracted from the current density operators (\ref{eq:current_densities}):
\begin{equation}
    \begin{gathered} \includegraphics[scale=0.2]{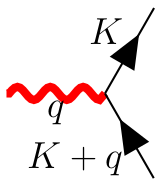} \end{gathered} = -i e_0 j_{\mu} = - i e_0 \Gamma^{(0)}_{\mu} \hspace{3pt}, \text{ with} \hspace{7pt} \Gamma^{(0)}_{\mu} = j_{\mu}.
    \label{eq:coupling_external_field}  
\end{equation}
In this work, we will use black wavy lines for Coulomb photons, whereas red ones will be reserved for external photons. Following \cite{Ahn2017}, let us define a $4 \times 4$ matrix $\mathcal{F}_{\mu \nu}(\varphi_{k}) = \text{diag}[1,\cos(\varphi_{k}),\sin(\varphi_{k}),1]$ containing the information about the geometric factors arising from the coupling to the external field (\ref{eq:coupling_external_field}). Then, the external vertex can be related to that of graphene \cite{Teber2018} as $\Gamma^{(0)}_{\mu} = \mathcal{F}_{\mu \nu}(\varphi_{k}) \Gamma^{(0)gr}_{\nu}$, where one has to understand $\Gamma^{(0)gr}_{x} = \Gamma^{(0)gr}_{y} \equiv \Gamma^{(0)gr}_{1} = \tau_x$ and $\Gamma^{(0)gr}_{z} \equiv \Gamma^{(0)gr}_{2} = \tau_y$. 

On the other hand, if we denote the counterterm insertions by $\begin{gathered} \includegraphics[scale=0.2]{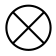} \end{gathered}$, the counterterm diagrams are:
\begin{gather}
\begin{aligned}
        \begin{gathered} \includegraphics[scale=0.2]{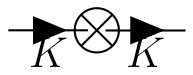} \end{gathered} = i [\delta_{\psi} \tau_0 \omega - (\delta_{\psi}+\delta_v) v_R \boldsymbol{\tau} \cdot \boldsymbol{k}],
\end{aligned} \\
\begin{aligned}
        \begin{gathered} \includegraphics[scale=0.2]{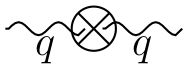} \end{gathered} = i \frac{1}{2} \delta_A \hspace{5pt} , \hspace{7pt} \begin{gathered} \includegraphics[scale=0.2]{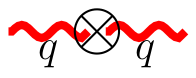} \end{gathered} = i \frac{1}{2} \delta_{A^{ext}_{\mu}} g_{\mu \nu},
    \label{eq:counterterm_field}
\end{aligned} \\
\begin{aligned}
        \begin{gathered} \includegraphics[scale=0.2]{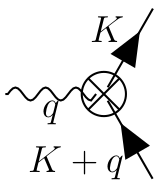} \end{gathered} =  - i \delta_{\text{Coul}} g_0 \tau_0 \hspace{2pt} , \hspace{-3pt} \begin{gathered} \includegraphics[scale=0.2]{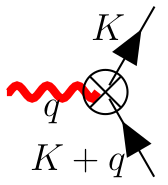} \end{gathered} =  - i \delta_{\Gamma_{\mu}} e_0 \Gamma^{(0)}_{\mu}.
        \label{eq:vertex_counterterm}
\end{aligned}
\end{gather}
Here, $g_{\mu \nu} = \text{diag}(+,-,-,-)$ is the spacetime metric.

Let us mention that the Coulomb field vertex and counterterms equal those of the time component of the external field by replacing $e_0 \leftrightarrow g_0$, a property we will take advantage of later.


Our task is to compute the (renormalized) polarization tensor to lowest order in the Coulomb interaction, which diagrammatically amounts to computing:
\begin{equation}
    \Pi_{\mu \nu}^{R}(q) = \Pi_{\mu \nu}^{R(0)}(q) + \Pi_{\mu \nu}^{R(1)}(q) + \mathcal{O}(e_R^6),
\end{equation}
with $\Pi_{\mu \nu}^{R(0)}$ the renormalized noninteracting polarization: 
\begin{align}
\begin{split}
    \Pi_{\mu \nu}^{R(0)}(q) & = \Pi_{\mu \nu}^{(0)}(q) + \Pi_{\mu \nu}^{c(0)}(q) = \\
    & \hspace{-20pt} = \begin{gathered} \includegraphics[scale=0.2]{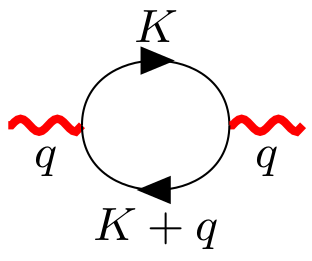} \end{gathered} + \begin{gathered} \includegraphics[scale=0.2]{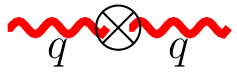} \end{gathered} \hspace{2pt} [\text{at } \mathcal{O}(e_R^2)],
\end{split}
\label{eq:one-loop_polarization}
\end{align}
and $\Pi_{\mu \nu}^{R(1)}$ the renormalized first interaction correction to the polarization:
\begin{equation}
    \Pi_{\mu \nu}^{R(1)}(q) = 2 \Pi_{\mu \nu}^{R(1a)}(q) + \Pi_{\mu \nu}^{R(1b)}(q) + \Pi_{\mu \nu}^{c(1)}(q),
     \label{eq:2-loop_polarization_diagrams_a}
\end{equation}
where $\Pi_{\mu \nu}^{R(1a)}$ is the renormalized self-energy correction:
\begin{equation}
\begin{split}
    \Pi_{\mu \nu}^{R(1a)}(q) & = \Pi_{\mu \nu}^{(1a)}(q) + \Pi_{\mu \nu}^{c(1a)}(q) = \\
    & = \begin{gathered} \includegraphics[scale=0.2]{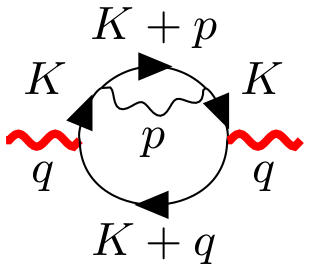} \end{gathered} + \begin{gathered} \includegraphics[scale=0.2]{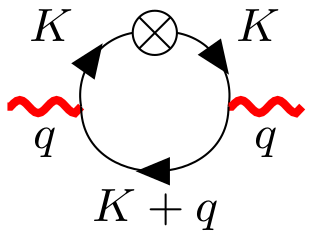} \end{gathered},
\end{split}
\label{eq:2-loop_polarization_self-energy_correction}
\end{equation} 
$\Pi_{\mu \nu}^{R(1b)}$ is the renormalized vertex correction:
\begin{equation}
\begin{split}
    \Pi_{\mu \nu}^{R(1b)}(q) & = \Pi_{\mu \nu}^{(1b)}(q) + 2 \Pi_{\mu \nu}^{c(1b)}(q) = \\
    & \hspace{-20pt} = \begin{gathered} \includegraphics[scale=0.2]{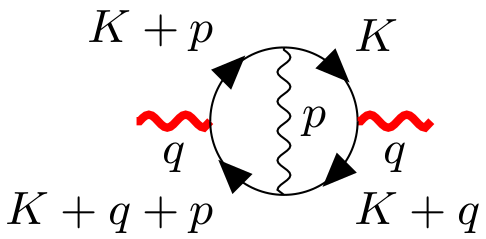} \end{gathered} \hspace{-10pt} +  2 \hspace{1pt} \begin{gathered} \includegraphics[scale=0.2]{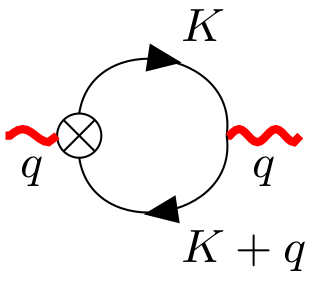} \end{gathered},
\end{split}
\label{eq:2-loop_polarization_vertex_correction}
\end{equation}
and $\Pi_{\mu \nu}^{c(1)}$ is the $\mathcal{O}(e_R^4)$ global counterterm:
\begin{equation}
    \Pi_{\mu \nu}^{c(1)}(q) = \begin{gathered} \includegraphics[scale=0.2]{counterterm_polarization_0.png} \end{gathered} \hspace{2pt} [\text{at } \mathcal{O}(e_R^4)].
    \label{eq:2-loop_polarization_diagrams_d}
\end{equation}
Note that the factors of 2 that appear in some diagrams arise from the number of equivalent forms the diagram can be written. The self-energy and vertex corrections to the polarization have subdiagrams corresponding to the one-loop electron self-energy $\Sigma^{(1)}(k)$ and the one-loop dressed vertex $\Lambda^{(1)}_{\mu}(k,q)$, respectively:
\begin{align}
\begin{split}
    -i \Sigma^{(1)}(k) = \begin{gathered} \includegraphics[scale=0.2]{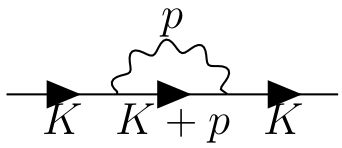} \end{gathered},
    \label{diag:electron_self-energy}
\end{split}\\
\begin{split}
    -i \Lambda^{(1)}_{\mu}(k,q) = \begin{gathered} \includegraphics[scale=0.2]{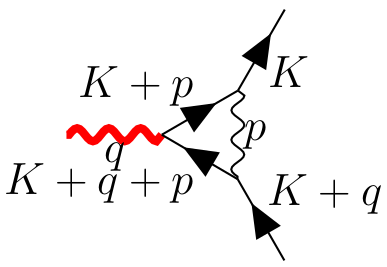} \end{gathered}.
    \label{eq:one-loop_vertex}
\end{split}    
\end{align}
Therefore, it will be useful to compute them before calculating the two-loop polarization diagrams.





\subsection{One-loop polarization and noninteracting conductivity}



Using the shorthand notation $\int_K$ for the $\tilde{d}$-dimensional integral $\int d^{\tilde{d}} K/(2 \pi)^{\tilde{d}}$, the one-loop polarization tensor reads:
\begin{equation}
\begin{split}
    i \Pi_{\mu \nu}^{(0)}(q) = - \int_K \text{Tr} & \left[ (-i e_0 \Gamma_{\nu}^{(0)}) S_R^{(0)}(K+q) \right. \\[-3pt] 
    & \hspace{5pt} \left.  (-i e_0 \Gamma_{\mu}^{(0)}) S_R^{(0)}(K) \right].
\end{split}
\end{equation}
In order to write the electron propagator of the sum of momenta, according to the definition (\ref{eq:free_electron_propagator}), we need $(\boldsymbol{K}+\boldsymbol{q})_r$. It will be useful to perform a change of coordinates by a rotation in the $xy$-plane to a new set of cartesian coordinates $(k_{\parallel},k_{\perp},k_z)$, with $k_{\parallel}$ and $k_{\perp}$ parallel and perpendicular to the projection of $\boldsymbol{k}$ in the $xy$-plane. These new coordinates are related to the old ones by $k_{\parallel} \equiv k_r$, $k_{\perp}=0$, $q_{\parallel} \equiv q_r \cos(\varphi_{qk})$ and $q_{\perp} \equiv q_r \sin(\varphi_{qk})$, where $\varphi_{qk}=\varphi_q-\varphi_k$ is the azimuthal angle between $\boldsymbol{q}$ and $\boldsymbol{k}$, i.e., the angle between the projections of $\boldsymbol{q}$ and $\boldsymbol{k}$ on the $xy$-plane. We also define the wave vector $q' = (\Omega,\boldsymbol{q'})$, with $q'_{\parallel}=q_{\parallel}=q_r \cos(\varphi_{qk})$, $q'_{\perp}=0$, and $q'_z=q_z$, i.e., the projection of $q$ on the $(\parallel z)$-plane. Now, assuming that $k_r,q_r \ll k_0$, which is consistent with our low-energy linear approximation of the dispersion, we can approximate $(\boldsymbol{K}+\boldsymbol{q})_r - k_0 = \sqrt{K_r^2 + 2 K_r q_r \cos(\varphi_{qk}) + q_r^2} -k_0 = k_{\parallel} + q'_{\parallel} + \mathcal{O}(1/k_0)$. The electron propagator of the sum of momenta may thus be written as:%
\small%
\begin{equation}
\begin{split}
    & \hspace{-3pt} S_R^{(0)}(K+q) = \\
    & \hspace{-3pt} = \hspace{-1pt} i \frac{\tau_0(\omega \hspace{-1pt}+\hspace{-1pt} \Omega) \hspace{-1pt}+\hspace{-1pt} v_R \{ \tau_x [(\boldsymbol{K}\hspace{-1pt}+\hspace{-1pt}\boldsymbol{q})_r \hspace{-1pt}-\hspace{-1pt} k_0] \hspace{-1pt}+\hspace{-1pt} \tau_y (\boldsymbol{K}\hspace{-1pt}+\hspace{-1pt}\boldsymbol{q})_z \} }{(\omega \hspace{-1pt}+\hspace{-1pt} \Omega)^2 \hspace{-1pt}-\hspace{-1pt} v_R^2 \{ [(\boldsymbol{K}\hspace{-1pt}+\hspace{-1pt}\boldsymbol{q})_r \hspace{-1pt}-\hspace{-1pt} k_0]^2 + (\boldsymbol{K}\hspace{-1pt}+\hspace{-1pt}\boldsymbol{q})_z^2 \} } \hspace{-1pt} \simeq  \\
    & \hspace{-3pt} \simeq i \frac{\tau_0(\omega + \Omega) + v_R \boldsymbol{\tau} \cdot (\boldsymbol{k}+\boldsymbol{q'})}{(\omega + \Omega)^2 - v_R^2 (\boldsymbol{k}+\boldsymbol{q'})^2 } \equiv S^{(0)gr}_R(k+q'),
    \label{eq:electron_propagator_aprox_graphene}
\end{split}
\end{equation}
\normalsize%
where $S^{(0)gr}_R(k)$ is the free fermion propagator of graphene in the coordinates $1 \equiv \parallel$ and $2 \equiv z$. Let us highlight that $q_{\perp}$ does not appear in the propagator at the zeroth order in the expansion in $1/k_0$, which is the level of approximation we are considering for computing the optical conductivity. In other words, remaining at the leading order in the $1/k_0$ expansion amounts to approximating, around each point of the line node, the 3D energy dispersion to an effectively 2D one only dispersing over the radial ($\parallel$) and $z$ directions and not over the tangent ($\perp$) direction. This reflects the analogy between each point of the line node and a 2D Dirac cone, with the two dispersing directions being $1 \equiv \parallel$ and $2 \equiv z$. 



Now, we change to the toroidal coordinates $(k_r \equiv k_{\parallel},\varphi_k,k_z)$ defined in Sec. \ref{model} and, as we anticipated, we approximate the integration in $k_r$ to be from $-\infty$ to $+\infty$, which is justified by the fact that our linear model is only valid up to momenta of the order of $k_0$, which acts as the UV-cutoff $\Lambda$ of our theory, $\Lambda \sim k_0$, and can thus be taken to infinity if we are interested in the physics at sufficiently smaller energies without affecting the physical results. Then, indicating explicitly the integration over the azimuthal angle and using that $\Gamma^{(0)}_{\mu} = \mathcal{F}_{\mu \nu}(\varphi_{k}) \Gamma^{(0)gr}_{\nu}$, the noninteracting polarization to lowest order in $1/k_0$ can be written as \cite{Ahn2017}:
\begin{equation}
    \Pi_{\mu \nu}^{(0)}(q) = k_0 \int \frac{d \varphi_{k}}{2 \pi} \mathcal{F}_{\mu \alpha}(\varphi_{k}) \mathcal{F}_{\nu \beta}(\varphi_{k}) \Pi_{\alpha \beta}^{(0)gr}(q'),
    \label{eq:polarization_NLSM-graphene_nonint}
\end{equation}
where $\Pi_{\mu \nu}^{(0)gr}$ is the polarization of a \textit{single spinless} Dirac cone of graphene, which is finite and has been calculated following the same steps as \cite{Teber2014,Teber2018}: first performing the trace by using the properties of the Pauli matrices, then Wick rotating to imaginary frequency to calculate the frequency integral and finally computing the integral in $D=2-2 \epsilon$ dimensions by means of the usual techniques \cite{schwartz2014quantum}. In fact, instead of calculating all the components $\Pi_{\mu \nu}^{(0)gr}$, one only needs to compute the density response $\Pi_{00}^{gr}$ and the trace $\Pi_{\mu \mu}^{gr}$, since with these two quantities the longitudinal $\Pi_L^{gr}$ and transverse $\Pi_T^{gr}$ polarizations are completely determined, and so are all the $\Pi_{i j}^{gr}$ components via the relation $\Pi_{ij}^{gr}=(q_iq_j/|\boldsymbol{q}|^2)\Pi_L^{gr} + (\delta_{ij}-q_iq_j/|\boldsymbol{q}|^2)\Pi_T^{gr}$ \cite{giuliani2005quantum,gonccalves2016introduction}. Let us note that the $\varphi_k$-dependence in the NLSM case comes both from the geometrical projection factor $\mathcal{F}_{\mu \alpha}(\varphi_{k})$ and from the momentum $q'$ \cite{Rhim2016} via $q'_{\parallel} = q_r \cos(\varphi_{qk})$. 


The finiteness of the noninteracting graphene polarization implies that of the NLSM $\Pi_{\mu \nu}^{(0)}$, and thus the counterterm diagram of (\ref{eq:one-loop_polarization}), $\Pi_{\mu \nu}^{c(0)} = (\delta_{A^{ext}_{\mu}}/2) g_{\mu \nu}$, vanishes, i.e., $\delta_{A^{ext}_{\mu}} = 0$, since the first correction to the external photon propagator $D_{\mu \nu}^{(1)}$ is already finite:
\begin{equation}
\begin{split}
& D_{\mu \nu}^{(1)}(q) = \begin{gathered} \includegraphics[scale=0.2]{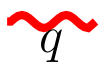} \end{gathered} + \begin{gathered} \includegraphics[scale=0.2]{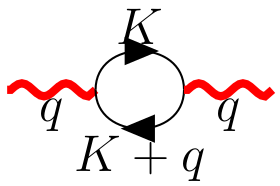} \end{gathered} + \begin{gathered} \includegraphics[scale=0.2]{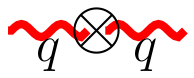} \end{gathered} = \\
& = \hspace{-2pt} D_{\mu \nu}^{(0)}(q) \hspace{-1pt} + \hspace{-1pt} D_{\mu \alpha}^{(0)}(q) i[\Pi_{\alpha \beta}^{(0)}(q) \hspace{-1pt} + \hspace{-1pt} \frac{\delta_{A^{ext}_{\alpha}}}{2} g_{\alpha \beta}] D_{\beta \nu}^{(0)}(q).\hspace{-2pt}
\end{split}
\end{equation}
Consequently, the renormalized one-loop polarization is $\Pi_{\mu \nu}^{R(0)} = \Pi_{\mu \nu}^{(0)}$. Moreover, since, as mentioned before, the Coulomb photon self-energy is $\Pi_{\text{Coul}}^{(0)} = (1/\varepsilon) \Pi_{00}^{(0)}$, the Coulomb field counterterm also vanishes, $\delta_{A} = 0$. Therefore, neither the external field nor the Coulomb field do renormalize at order $\mathcal{O}(e_R^2)$.

Now, using Eq. (\ref{eq:polarization_NLSM-graphene_nonint}), the conductivity $\sigma_{ij}^{(0)}(q) = -i (1/\Omega) \Pi_{ij}^{R(0)}(q)$ to zeroth order in the Coulomb interaction can be deduced to be:
\begin{equation}
    \sigma_{ij}^{(0)}(q) = k_0 \int \frac{d \varphi_{k}}{2 \pi} \mathcal{F}_{il}(\varphi_{k}) \mathcal{F}_{jm}(\varphi_{k}) \sigma_{lm}^{(0)gr}(q').
\end{equation}
In the long-wavelength limit $\boldsymbol{q}\rightarrow0$ and for the noninteracting case we are considering now, graphene conductivity is diagonal (as long as time-reversal symmetry is preserved), frequency-independent, and reads as $\sigma_{ij}^{(0)gr}(\Omega) = \delta_{ij} e_R^2 / 16$ \cite{Teber2014,Teber2018}, so the NLSM conductivity is:
\begin{equation}
	\sigma_{zz}^{(0)}(\Omega) = \sigma_0 \hspace{5pt} , \hspace{10pt} \sigma_{xx}^{(0)}(\Omega) = \sigma_{yy}^{(0)}(\Omega) = \frac{1}{2} \sigma_0,
	\label{eq:free_conductivity}
\end{equation}
where $\sigma_0 = k_0 e_R^2/16$, in agreement with \cite{Ahn2017}. Let us point out that in the noninteracting case it is straightforward to consider anisotropic Fermi velocities by a simple rescaling of momenta, which gives the following conductivities: $\sigma_{zz}^{(0)}(\Omega) = (v_z/v_r) \sigma_0$ and $\sigma_{xx}^{(0)}(\Omega) = \sigma_{yy}^{(0)}(\Omega) = (v_r/v_z) \sigma_0/2$, in agreement with \cite{Huh2016,Shao2019}. 


As a crosscheck for our results, we can verify that the identity $q_i q_j \Pi_{ij} = q_0^2 \Pi_{00}$, which stems from the transversality $q_{\mu} \Pi_{\mu \nu} = \Pi_{\mu \nu} q_{\nu} = 0$ of the polarization required by gauge invariance and total particle conservation \cite{schakel2008boulevard,giuliani2005quantum,gonccalves2016introduction}, is verified. In fact, using expression (\ref{eq:polarization_NLSM-graphene_nonint}) with $\Pi_{00}^{(0)gr}(\Omega,\boldsymbol{q'}\rightarrow0) = - e_R^2 |\boldsymbol{q'}|^2 / (16 i \Omega)$ \cite{Teber2018}, we arrive at the following NLSM density-density response:
\begin{equation}
	\Pi_{00}^{(0)}(\Omega,\boldsymbol{q}\rightarrow0) = - k_0 \frac{e_R^2}{16} \frac{1}{i \Omega} \left( \frac{1}{2} q_r^2 +q_z^2 \right).
	\label{eq:density_density_external}
\end{equation} 
The transversality condition can now be checked by substitution. Let us also mention that due to the anisotropy of our system (arising from the presence of the nodal loop independently from the Fermi velocities), the longitudinal conductivity $\sigma_L^{(0)} = \sigma_0 [(1/2)q_r^2+q_z^2]/|\boldsymbol{q}|^2=\sigma_0 [1+\cos^2(\theta)]/2$, which is the relevant one in the study of plasmons \cite{Rhim2016}, depends on the polar angle $\theta$ of the external wave vector ($q_r=|\boldsymbol{q}| \sin(\theta)$, $q_z=|\boldsymbol{q}| \cos(\theta)$).

\subsection{Electron self-energy and velocity renormalization}


The electron self-energy (\ref{diag:electron_self-energy}), using expression (\ref{eq:electron_propagator_aprox_graphene}) for the electron propagator, is:%
\begin{equation}
    \hspace{-3pt}-\hspace{-1pt}i \Sigma^{(1)}\hspace{-1pt}(k) \hspace{-2pt} = \hspace{-5pt} \int_p \hspace{-3pt}  (-i g_0 \tau_0) S_R^{(0)gr}\hspace{-2pt}(k+p') (-i g_0 \tau_0) V_R^{(0)}\hspace{-1pt}(p).\hspace{-3pt} 
\end{equation}%
Working in the $(k_{\parallel},k_{\perp},k_z)$-coordinates, the $p_{\perp}$-dependence only appears through the free photon propagator $V_R^{(0)}(p) = i / (p_{\parallel}^2 + p_{\perp}^2 +p_z^2)$. We can thus directly integrate it out to produce an effectively 2D photon propagator $V^{(0)gr}_R(p) = i/\left(2 \sqrt{p_{\parallel}^2+p_z^2}\right) \equiv i/\left(2 \sqrt{{p'_{\parallel}}^{2}+{p'_z}^{2}}\right) \equiv V^{(0)gr}_R(p')$, which is exactly the free photon propagator in graphene \cite{Teber2014,Teber2018} in the coordinates $1 \equiv \parallel$ and $2 \equiv z$. Therefore, the NLSM self-energy equals that of graphene [plus corrections $\mathcal{O}(1/k_0)$], $\Sigma^{(1)}(k) = \Sigma^{(1)gr}(k)$. Following the same calculations as \cite{Teber2014,Teber2018} (Wick rotating to imaginary frequency, performing the frequency integral and calculating the integral in $D=2-2 \epsilon$ dimensions), we finally arrive at:
\begin{subequations}
\begin{align}
    & \Sigma^{(1)}(k) = - \frac{\alpha_R}{8} e^{\gamma_E \epsilon} \left(\frac{\bar{\mu}^2}{|\boldsymbol{k}|^2} \right)^{\epsilon} G\left( \epsilon \right) (v_R \boldsymbol{\tau} \cdot \boldsymbol{k}) = \\
    & = - \frac{\alpha_R}{8} \left( \frac{1}{\epsilon} - \ln{\frac{|\boldsymbol{k}|^2}{\bar{\mu}^2}} + 4 \ln{2} \right) (v_R \boldsymbol{\tau} \cdot \boldsymbol{k}) + \mathcal{O}(\epsilon),\hspace{-2pt}
\end{align}
\end{subequations}
where we have taken the $\epsilon \rightarrow 0$ limit in the second equality and we have defined the renormalized fine-structure constant of our isotropic Fermi-velocity NLSM $\alpha_R = g_R^2 / (4 \pi v_R) = e_R^2 / (4 \pi \varepsilon v_R)$, which is the effective coupling constant of the Coulomb interaction, as in graphene. Furthermore, we have defined the function $G(\epsilon) = [\Gamma(1/2-\epsilon)]^2 \Gamma(\epsilon) / [\pi \Gamma(1-2 \epsilon)]$, where $\Gamma(x)$ is the gamma-function.


Although the (bare) self-energy diverges, we have to choose the appropriate counterterms so that the electron propagator $S_R^{(1)}(k)$ to first order in the coupling $\alpha_R$ remains finite: 
\begin{equation}
\begin{split}
    & S_R^{(1)}(K) = \begin{gathered} \includegraphics[scale=0.2]{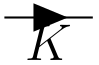} \end{gathered} + \begin{gathered} \includegraphics[scale=0.2]{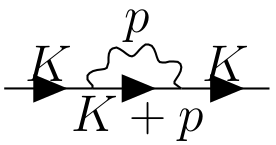} \end{gathered} + \begin{gathered} \includegraphics[scale=0.2]{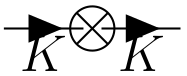} \end{gathered} = \\
    & = S_R^{(0)}(K) + S_R^{(0)}(K) i \Sigma^{(1)}(k) S_R^{(0)}(K) + \\ 
    & \hspace{10pt} + S_R^{(0)}(K) i [\delta_{\psi} \tau_0 \omega - (\delta_{\psi}+\delta_v) v_R \boldsymbol{\tau} \cdot \boldsymbol{k}] S_R^{(0)}(K).\hspace{-3pt}
\end{split}
\end{equation}
This implies, using the $\bar{\text{MS}}$ subtraction scheme and up to order $\mathcal{O}(\alpha_R) = \mathcal{O}(e_R^2)$, that:
\begin{equation}
    \delta_{\psi} = 0 \hspace{10pt} , \hspace{15pt} \delta_v = - \frac{1}{\epsilon} \frac{\alpha_R}{8},
\end{equation}
i.e., the wavefunction is not renormalized while the velocity is. As a crosscheck, we can compute the velocity $\beta$-function \cite{Teber2018} and compare it to the literature:
\begin{equation}
    \beta_v \equiv \bar{\mu} \frac{\partial v_R}{\partial \bar{\mu}} = 2 v_R \delta_v \epsilon = - v_R \frac{\alpha_R}{4}.
\label{eq:beta_RG_v}
\end{equation}
Indeed, this negative $\beta$ function, which means that velocity grows in the infrared (i.e., with decreasing frequency) as in graphene \cite{Gonzalez1994}, coincides with the one provided in \cite{Huh2016} for the isotropic-Fermi-velocity NLSM we are considering here. The Coulomb coupling $\beta$-function to $\mathcal{O}(\alpha_R^2)$ can also be easily derived from its definition and $\beta_v$:
\begin{equation}
\beta_{\alpha} \equiv \bar{\mu} \frac{\partial \alpha_R}{\partial \bar{\mu}} = \frac{\alpha_R^2}{4}.
\label{eq:beta_RG_alpha}
\end{equation}
As already computed by \cite{Huh2016}, the Coulomb coupling flows to weak coupling (i.e., decreases in the infrared), determining its marginally irrelevant character.

\subsection{One-loop interaction vertex and Ward identities}

We now proceed to calculate the one-loop correction to the interaction vertex between the fermionic and the external electromagnetic fields, $\Gamma_{\mu}^{(1)} = \Gamma_{\mu}^{(0)} + \Lambda_{\mu}^{(1)}$, where:
\begin{equation}
\begin{split}
    -i e_0 & \Lambda_{\mu}^{(1)}(k,q) = \int_p (-i g_0 \tau_0) S_R^{(0)gr}(k+p'+q') \\ 
    & \hspace{10pt} (-i e_0 \Gamma_{\mu}^{(0)}) S_R^{(0)gr}(k+p') (-i g_0 \tau_0) V_R^{(0)}(p).\hspace{-2pt}
\end{split}
\end{equation}
As we did in the calculation of the self-energy, let us work in the $(k_{\parallel},k_{\perp},k_z)$-coordinates. 
As in the former case, the only $p_{\perp}$ dependence occurs at the photon propagator $V_R^{(0)}(p)$. Therefore, we can integrate in $p_{\perp}$ to obtain $V_{R}^{(0)gr}(p')$. Consequently, we can again take advantage of the vertex correction in graphene $\Lambda_{\mu}^{(1)gr}$ and write:
\begin{equation}
    \Lambda_{\mu}^{(1)}(k,q) = \mathcal{F}_{\mu \nu}(\varphi_k) \Lambda_{\nu}^{(1)gr}(k,q'),
    \label{eq:vertex_correction_NLSM_graphene}
\end{equation}
where $q'_{\parallel}=q_r \cos(\varphi_{qk})$ and $q'_z=q_z$.

Before continuing to the two-loop calculations, let us make the following remark. Gauge invariance of the theory imposes some constrictions, the so-called Ward identities \cite{Teber2018,Gonzalez1994,schwartz2014quantum}. One of them is that the renormalization constant of the Coulomb interaction vertex $Z_{\text{Coul}}$, or equivalently of the time component of the external field vertex $Z_{\Gamma_0}$, must equal that of the time component of the kinetic term, $Z_{\psi}$, i.e., $Z_{\text{Coul}} = Z_{\Gamma_0} = Z_{\psi}$. Since $Z_{\text{Coul}} = Z_{\psi} Z_{e} Z_{A}^{1/2}$ and we have previously determined that neither the wavefunction nor the gauge field do renormalize at order $\mathcal{O}(e_R^2)$ in the NLSM, $Z_{\psi}=Z_{A}=1 + \mathcal{O}(e_R^4)$, then the Ward identity implies that the charge is not renormalized either, $Z_e=1 + \mathcal{O}(e_R^4)$, a well known property in graphene \cite{Teber2018,Gonzalez1994}. Other Ward identity consists of the equality of the spatial components of the external field vertex and the kinetic term, $Z_{\Gamma_{i}} = Z_{\psi} Z_v$, which implies that $\delta_{\Gamma_{i}} = \delta_v + \mathcal{O}(e_R^4)$ in the NLSM.


We can indeed check that the Ward identities are verified. Firstly, $\Lambda_{\text{Coul}}^{(1)} = \Lambda_{0}^{(1)}$ are finite for the NLSM due to the finiteness of $\Lambda_{\text{Coul}}^{(1)gr}$ for graphene \cite{Teber2018}, which implies that the vertex counterterm vanishes, $\delta_{\text{Coul}}=\delta_{\Gamma_0} = 0+\mathcal{O}(e_R^4)$, or equivalently $Z_{\text{Coul}} = Z_{\Gamma_0} = 1+\mathcal{O}(e_R^4)$, which proves the first Ward identity since $Z_{\psi} = 1 +\mathcal{O}(e_R^4)$. Secondly, using the relation (\ref{eq:vertex_correction_NLSM_graphene}) and the divergent part of the spatial component of the dressed vertex in graphene \cite{Teber2018}, $\Lambda_i^{(1)gr}(k,q')=\alpha_R/(8 \epsilon) v_R \tau_i + \mathcal{O}(\epsilon^0)$, we can write the vertex correction for the NLSM as:
\begin{equation}
\Lambda_i^{(1)}(k,q)= \frac{\alpha_R}{8 \epsilon} j_i(\varphi_k) + \mathcal{O}(\epsilon^0) = \frac{\alpha_R}{8 \epsilon} \Gamma_i^{(0)} + \mathcal{O}(\epsilon^0). \hspace{-2pt}  
\end{equation}
Considering also the spatial components of the counterterm diagram (\ref{eq:vertex_counterterm}), the renormalized first order external interaction vertex is therefore:
\small%
\begin{equation}
    \Gamma_{i}^{R(1)} \hspace{-3pt} = \hspace{-1pt} \Gamma_{i}^{(0)} \hspace{-2pt} + \hspace{-1pt} \Lambda_{i}^{(1)} \hspace{-2pt} + \hspace{-1pt} \delta_{\Gamma_i} \Gamma_i^{(0)} \hspace{-2pt} = \hspace{-1pt} \Gamma_{i}^{(0)} \left[1 \hspace{-1pt} + \hspace{-1pt} \frac{\alpha_R}{8 \epsilon} \hspace{-1pt} + \hspace{-1pt} \delta_{\Gamma_i}\right] \hspace{-1pt} + \hspace{-1pt} \mathcal{O}(\epsilon^0). \hspace{-2pt}
\end{equation}
\normalsize%
Its finiteness implies that the spatial vertex counterterm is $\delta_{\Gamma_i}=-\alpha_R/(8 \epsilon)\equiv \delta_v$, which is exactly the requirement imposed by the second Ward identity.

\subsection{Two-loop polarization and interaction corrections to conductivity}

The Feynman diagrams contributing to the first Coulomb interaction correction to the polarization tensor are sketched in expressions (\ref{eq:2-loop_polarization_diagrams_a}-\ref{eq:2-loop_polarization_diagrams_d}). They read as:
\begin{equation}
\begin{split}
    & \hspace{-1pt} i 2 \Pi_{\mu \nu}^{R(1a)}(q) = i 2 \left[\Pi_{\mu \nu}^{(1a)}(q) + \Pi_{\mu \nu}^{c(1a)}(q) \right] = \\
    & \hspace{-1pt} = -2 \int_k  \text{Tr} \left\{ (-i e_0 \Gamma_{\nu}^{(0)}) S_R^{(0)gr}(k+q') (-i e_0 \Gamma_{\mu}^{(0)}) \right. \hspace{-1pt}  \hspace{-2pt} \\ \hspace{-1pt}
    & \hspace{-1pt} \hspace{-2pt} \left. S_R^{(0)gr}(k) \left[-i \left(\Sigma^{(1)}(k) + \delta_v v_R \boldsymbol{\tau} \cdot \boldsymbol{k} \right) \right] S_R^{(0)gr}(k) \right\}, \hspace{-2pt}
\end{split}
\end{equation}
and
\begin{equation}
\begin{split} 
    & i \Pi_{\mu \nu}^{R(1b)}(q) = i \left[\Pi_{\mu \nu}^{(1b)}(q)+ 2 \Pi_{\mu \nu}^{c(1b)}(q) \right] = \\
    & = - \int_k \text{Tr} \left\{ (-i e_0 \Gamma_{\nu}^{(0)}) S_R^{(0)gr}(k+q') \right. \\ 
    & \hspace{11pt} \left. \left[-i e_0 \left( \Lambda_{\mu}^{(1)}(k,q') + 2\delta_{\Gamma_{\mu}} \Gamma_{\mu}^{(0)} \right) \right] S_R^{(0)gr}(k) \right\}. \hspace{-2pt}
\end{split}
\end{equation}

Using the analogies between NLSMs and graphene, in particular $\Gamma_{\mu}^{(0)}=\mathcal{F}_{\mu \alpha} \Gamma_{\mu}^{(0)gr}$, $\Sigma^{(1)} = \Sigma^{(1)gr}$, $\delta_v = \delta_v^{gr}$, $\Lambda_{\mu}^{(1)} = \mathcal{F}_{\mu \alpha} \Lambda_{\alpha}^{(1)gr}$ and $\delta_{\Gamma_{\mu}}=\delta_{\Gamma_{\mu}}^{gr}$, and working in toroidal coordinates $(k_r \equiv k_{\parallel},\varphi_k,k_z)$, we can write that:
\begin{equation}
    \Pi_{\mu \nu}^{R(1x)}(q) = k_0 \int_{\varphi}  \mathcal{F}_{\mu \alpha}(\varphi_k) \mathcal{F}_{\nu \beta}(\varphi_k)  \Pi_{\alpha \beta}^{R(1x)gr}(q'), 
\end{equation}
where $x=a,b$ and we have used the shorthand notation $\int_{\varphi}$ for the angular integration $\int d \varphi_k/(2 \pi)$. The calculation of $\Pi_{\mu \nu}^{R(1x)gr}$ follows the same lines as in the non-interacting case: first performing the trace, then Wick rotating to imaginary frequency to perform the frequency integral and eventually using the master integrals of \cite{Teber2014,Kotikov2019} in $D=2-2 \epsilon$ dimensions. And once again, the easiest way to do it is to take advantage of the decomposition of the graphene polarization in longitudinal and transverse and calculate the time component and the trace.  As before, we have followed the same procedure as \cite{Teber2018}, obtaining the same intermediate as well as final results, so we refer to that paper for further details of the calculation. Indeed, in these interaction corrections a little subtlety has been done in a different but analogous way to \cite{Teber2018}. Here, we use conventional renormalization (see, e.g., \cite{schwartz2014quantum}) and draw explicitly the counterterm diagrams that substract the subdivergences of the self-energy and vertex corrections [see the second diagrams of (\ref{eq:2-loop_polarization_self-energy_correction}) and (\ref{eq:2-loop_polarization_vertex_correction})]. On the other hand, Ref. \cite{Teber2018} applies the Bogoliubov-Parasiuk-Hepp-Zimmermann (BPHZ) renormalization prescription, where the subdivergences are substracted directly without explicitly drawing the counterterm diagrams (see, e.g., \cite{collins1985renormalization}). Nevertheless, both procedures are completely equivalent.


Since $\Pi_{\mu \nu}^{R(1x)gr}$ turns out to be finite in graphene \cite{Teber2018}, the corresponding $\Pi_{\mu \nu}^{R(1x)}$ for the NLSM is also finite, and thus there is no global divergence. Therefore, the global $\mathcal{O}(e_R^4)$ counterterm (\ref{eq:2-loop_polarization_diagrams_d}) vanishes, $\Pi_{\mu \nu}^{c(1)} = 0$, i.e., $\delta_A = \delta_{A_{\mu}^{ext}} = 0 + \mathcal{O}(e_R^6)$, as in graphene. Then,
\begin{equation}
\begin{split}
	\Pi_{\mu \nu}^{R(1)}&(q) = 2 \Pi_{\mu \nu}^{R(1a)}(q) + \Pi_{\mu \nu}^{R(1b)}(q) = \\
	& = k_0 \int \frac{d \varphi_k}{2 \pi} \mathcal{F}_{\mu \alpha}(\varphi_k) \mathcal{F}_{\nu \beta}(\varphi_k)  \Pi_{\alpha \beta}^{R(1)gr}(q').
\end{split}
\end{equation}



The first interaction correction to the conductivity in our NLSM $\sigma_{ij}^{(1)}$ is thus easily computed from the corresponding one in a \textit{single} Dirac cone of graphene $\sigma_{ij}^{(1)gr}$. We have obtained it to be:
\begin{equation}
	\sigma_{ij}^{(1)gr}= \frac{e_R^2}{16} C_2 \alpha_R \delta_{ij},
\end{equation}
where the first-order coefficient is $C_2 = (19-6 \pi)/12 \simeq 0.013$, the value first obtained by \cite{Mishchenko2008} and the more accepted up to date. Consequently, for the NLSM, performing the integral over the azimuthal angle:
\begin{equation}
	\sigma_{zz}^{(1)}(\Omega) = 2 \sigma_{xx}^{(1)}(\Omega) = 2 \sigma_{yy}^{(1)}(\Omega) = \sigma_0 C_2 \alpha_R .
	\label{eq:interacting_conductivity}
\end{equation}
The full expression of the optical conductivity presented in the main text can then be deduced from expressions (\ref{eq:free_conductivity}) and (\ref{eq:interacting_conductivity}) after reinstating the $\hbar$ factors where appropriate.





\bibliographystyle{apsrev4-1}
\bibliography{Draft_OC_NLSMs_Munoz-Segovia_and_Cortijo.bib}


\end{document}